\documentclass[journal]{IEEEtran}

\usepackage{amsmath}
\usepackage{amssymb}
\usepackage{balance}
\usepackage{bm}
\usepackage{cite}
\usepackage{color}
\usepackage[T1]{fontenc}
\usepackage{graphicx}
\usepackage[utf8]{inputenc}
\usepackage{mathtools}
\usepackage{multirow}
\usepackage{siunitx}
\usepackage{threeparttable}

\graphicspath{{./}{../figures/}}

\newtheorem{theorem}{Proposition}

\newlength{\mylength}
\newlength{\myvspace}

\begin{document}

\newcommand{\myauthor}{Author 1, Author 2, and Author 3}
\newcommand{\mytitle}{Energy-Efficient Time Synchronization Based on
  Asynchronous Source Clock Frequency Recovery and Reverse Two-Way Message
  Exchanges in Wireless Sensor Networks}

\title{\LARGE \mytitle}

\hypersetup{%
  pdfauthor={\myauthor},%
  pdftitle={\LARGE \mytitle},%
  pdfkeywords={},%
  pdfsubject={},%
}

\author{%
  Kyeong~Soo~Kim,~\IEEEmembership{Member, IEEE,}~%
  Sanghyuk~Lee,~and~%
  Eng~Gee~Lim,~\IEEEmembership{Member, IEEE}%
  \thanks{%
    This work was supported in part by Xi'an Jiaotong-Liverpool University
    Centre for Smart Grid and Information Convergence (CeSGIC) and Research
    Development Fund (under Grant RDF-14-01-25). An earlier version of this
    paper was presented in part at ICISCA 2015, Kuala Lumpur, Malaysia, June
    2015, and ISAE 2015, Busan, Korea, October 2015.}%
  \thanks{%
    K. S. Kim, S. Lee, and E. G. Lim are with the Department of Electrical and
    Electronic Engineering, Xi'an Jiaotong-Liverpool University, Suzhou 215123,
    Jiangsu Province, P. R. China (e-mail: \{Kyeongsoo.Kim, Sanghyuk.Lee,
    Enggee.Lim\}@xjtlu.edu.cn).%
  }%
}%


\maketitle

\begin{abstract}
  We consider energy-efficient time synchronization in a wireless sensor network
  where a head node (i.e., a gateway between wired and wireless networks and a
  center of data fusion) is equipped with a powerful processor and supplied
  power from outlet, and sensor nodes (i.e., nodes measuring data and connected
  only through wireless channels) are limited in processing and
  battery-powered. It is this \textit{asymmetry} that our study focuses on;
  unlike most existing schemes to save the power of all network nodes, we
  concentrate on battery-powered sensor nodes in minimizing energy consumption
  for time synchronization. We present a time synchronization scheme based on
  asynchronous source clock frequency recovery and reverse two-way message
  exchanges combined with measurement data report messages, where we minimize
  the number of message transmissions from sensor nodes, and carry out the
  performance analysis of the estimation of both measurement time and clock
  frequency with lower bounds for the latter. Simulation results verify that the
  proposed scheme outperforms the schemes based on conventional two-way message
  exchanges with and without clock frequency recovery in terms of the accuracy
  of measurement time estimation and the number of message transmissions and
  receptions at sensor nodes as an indirect measure of energy efficiency.
\end{abstract}

\begin{IEEEkeywords}
  Time synchronization, energy efficiency, source clock frequency recovery,
  two-way message exchanges, wireless sensor networks.
\end{IEEEkeywords}

\IEEEpeerreviewmaketitle

\section{Introduction}
\label{sec:introduction}
\IEEEPARstart{A}{} common time frame among network nodes in a wireless sensor
network (WSN) is critical for their carrying out important operations like
fusing data from different sensor nodes, time-based channel sharing and media
access control (MAC) protocols, and coordinated sleep wake-up node scheduling
mechanisms \cite{wu11:_clock_synch_wirel_sensor_networ}. In a typical WSN, a
head/master node (i.e., a base station that serves as a gateway between wired
and wireless networks and a center for fusion of data from distributed sensors)
is equipped with a powerful processor and supplied power from outlet, while
sensor/slave nodes (i.e., nodes measuring data with sensors and connected to a
WSN only through wireless channels) are limited in processing and
battery-powered. It is this \emph{asymmetry} that our study focuses on; unlike
most existing schemes trying to save the power of all network nodes (e.g.,
\cite{akhlaq13:_rtsp} and \cite{macii09:_power_wirel_sensor_networ}), we
concentrate on battery-powered sensor nodes, which are many in number, in
minimizing energy consumption for time synchronization.

In this paper we mean by \textit{time synchronization} a process of establishing
a common time frame with which the nodes in a network can operate one another,
whether their clocks are synchronized or not. By \textit{clock synchronization},
on the other hand, we mean a process of synchronizing each node clock to that of
a common reference node, typically through a logical clock that is a function of
a physical clock, which is one way of achieving time synchronization. Note that
without synchronizing node clocks, we can still provide a common time frame for
the operation of network nodes. For instance, in multi-hop extension of the
reference broadcast synchronization (RBS) algorithm \cite{elson02:_fine}, there
are multiple broadcast domains independently maintaining their own clocks. In
this case, the common time frame among multiple broadcast domains is provided
through time conversion at gateway nodes belonging to neighboring domains.

In the literature, time synchronization is formulated as the problem of
estimating clock parameters, often including node distances, in a pairwise
(distributed) or a network-wide (global) manner
\cite{trump01:_maxim,rajan11:_joint,chepuri13:_joint,Moon:99}. Few works,
however, have focused on the mode of operation (i.e., time vs. clock
synchronization discussed above) and the way of implementation related with
energy efficiency. To achieve better performance, several joint estimation
algorithms for both clock parameters and pair-wise distances have been proposed
(e.g., \cite{chepuri13:_joint} and \cite{rajan11:_joint}). Most of them are,
however, based on centralized and global offline algorithms, and the estimation
of parameters and the way of implementing clock/time synchronization together
with the delivery of the results of estimation are not explicitly
discussed. While joint estimation algorithms usually assume that all sensor
nodes are available in the beginning, recursive and online operation of
synchronization schemes are critical for WSN applications because there are
nodes who join later or resume their operations in the middle.

In this paper we present an energy-efficient time synchronization scheme based
on asynchronous source clock frequency recovery (SCFR) \cite{Kim:13-1} and
\textit{reverse} two-way message exchanges combined with measuring data report
messages, where we minimize the number of message transmissions and receptions
at sensor nodes, especially the number of message transmissions noting that the
energy for message transmission is typically higher than that for message
reception \cite{mainwaring02:_wirel_sensor_networ_habit_monit}. In the proposed
time synchronization scheme, only the frequency of a sensor node clock is
synchronized to that of the reference clock at the head node, but not its clock
offset; the proposed scheme is based on the idea of the separation of the clock
frequency\footnote{We use the terms ``clock frequency'' and ``clock skew''
  interchangeably in this paper.} estimation/compensation at sensor nodes and
the clock offset and delay estimation at the head node. For the clock frequency
recovery, each sensor node passively listens to any messages with timestamps
either broadcasted (e.g., beacons) or unicasted (e.g., control messages to a
specific node) from the head node and carries out asynchronous source clock
frequency recovery described in \cite{Kim:13-1}, which is basically one-way
clock frequency estimation. For the clock offset and delay estimation, a simple
two-way message exchange procedure
\cite{mills91:_inter,ganeriwal03:_timin_protoc_sensor_networ} is used but in a
reverse direction where the head node initiates the procedure and keeps track of
the offsets between its reference clock and sensor node clocks; also, instead of
dedicated, periodical synchronization message exchanges, we embed the
synchronization ``Response'' messages of the two-way message exchange procedure
in the measurement data report messages from sensor nodes in order to minimize
the number of message transmissions. In this way we can move most of time
synchronization operations to the head node and reduce the complexity and
thereby power consumption of sensor nodes for time synchronization. To carry out
usual WSN operations with the proposed time synchronization scheme, the head
node translates timestamp values based on clock offset information before
transmitting control messages to sensor nodes.

The major contributions of this work are three-fold: First, it provides a new
energy-efficient time synchronization scheme for asymmetric WSNs to minimize the
number of two-way message exchanges by cleverly combining one-way clock skew
estimation/compensation and reverse two-way message exchanges. Note that the
two-way message exchanges cannot be avoided in estimating both clock offset and
clock skew because it is impossible to separate the effects of clock offset and
propagation delay with one-way message exchanges alone
\cite{wu11:_clock_synch_wirel_sensor_networ,Kim:14-4}. The proposed scheme does
not only minimize the number of two-way message exchanges but also enable the
use of simple, low-complexity one-way clock skew estimation/compensation
algorithms to avoid the high complexity of advanced estimation algorithms taking
into nuisance parameters like mean and variance of the propagation
delay. Secondly, we formally describe the operations of and analyze the time
synchronization performance of the proposed scheme: By separately modeling times
of a hardware clock and a logical clock at a network node, we can capture the
dynamic nature of the proposed time synchronization scheme in the analysis,
which estimates clock parameters of a counterpart and updates its own logical
clock recursively. Thirdly, we also analyze the performance of measurement time
estimation based on both conventional and reverse two-way message exchanges. As
for clock skew estimation/compensation, we carry out a comparative analysis of
the performance of one-way and two-way maximum likelihood (ML) and ML-like
estimators and derive Cram\'{e}r-Rao lower bounds and lower bounds for ML and
ML-like estimators, respectively.

The rest of the paper is organized as follows: In
Section~\ref{sec:energy-effic-time}, we describe the proposed time
synchronization scheme with hardware and logical clock models; we also analyze
the effect of clock skew on measurement time estimation in both conventional and
reverse two-way message exchanges, present joint ML and ML-like one-way clock
skew estimators with their performance bounds, and discuss an extension to
multi-hop time synchronization through gateway
nodes. Section~\ref{sec:simulation-results} presents the results of simulations
for a comparative analysis of the performance of one-way and two-way clock skew
estimators for asynchronous SCFR and the investigation of the time
synchronization performance and the energy efficiency of the proposed scheme
compared to the schemes based on conventional two-way message exchanges with and
without SCFR. Section~\ref{sec:comp-relat-work} reviews the related work in
comparison to our work before concluding this paper in
Section~\ref{sec:conclusions}.

\section{Energy-Efficient Time Synchronization for Asymmetric WSNs}
\label{sec:energy-effic-time}
\begin{figure}[!tb]
  \begin{center}
    \includegraphics[angle=-90,width=.9\linewidth]{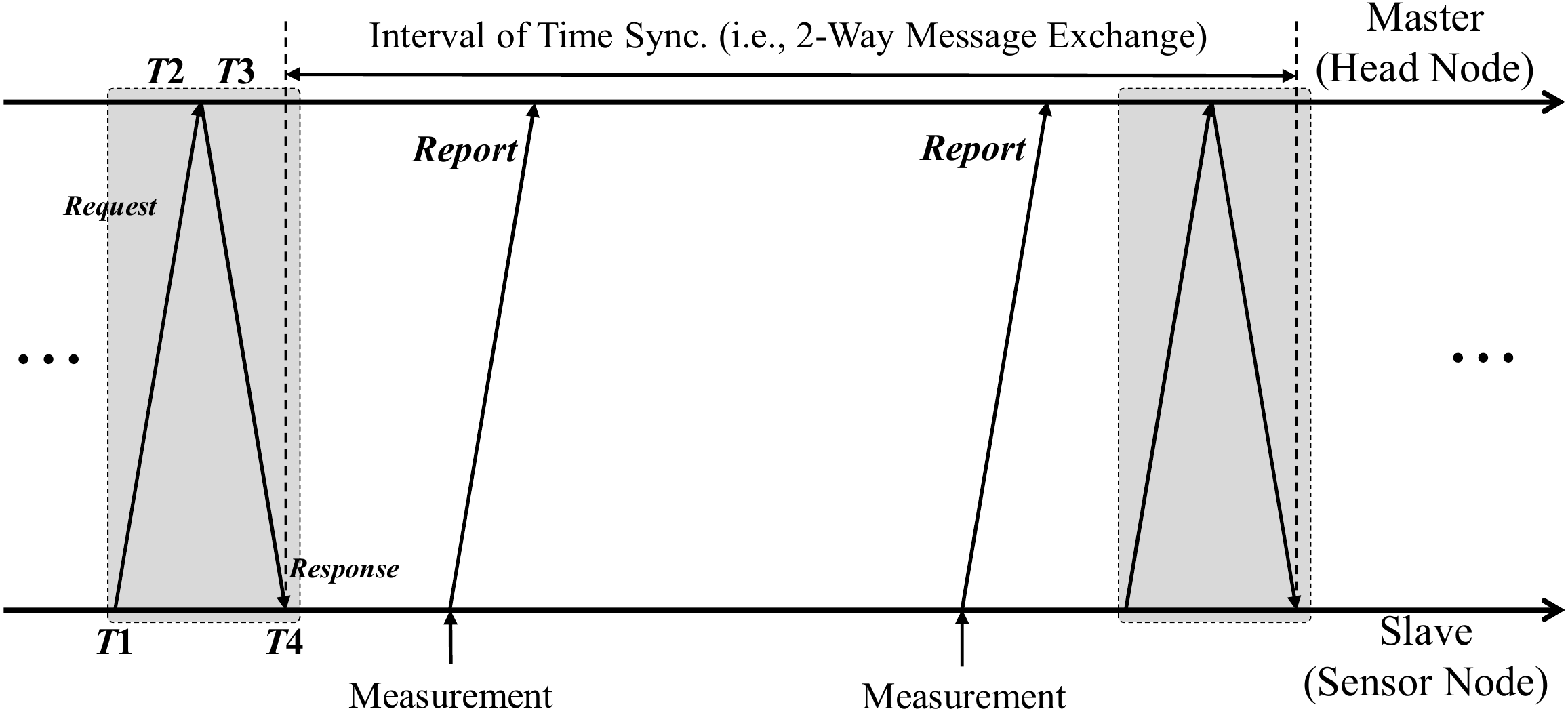}\\
    {\scriptsize (a)}\\
    \includegraphics[angle=-90,width=.9\linewidth]{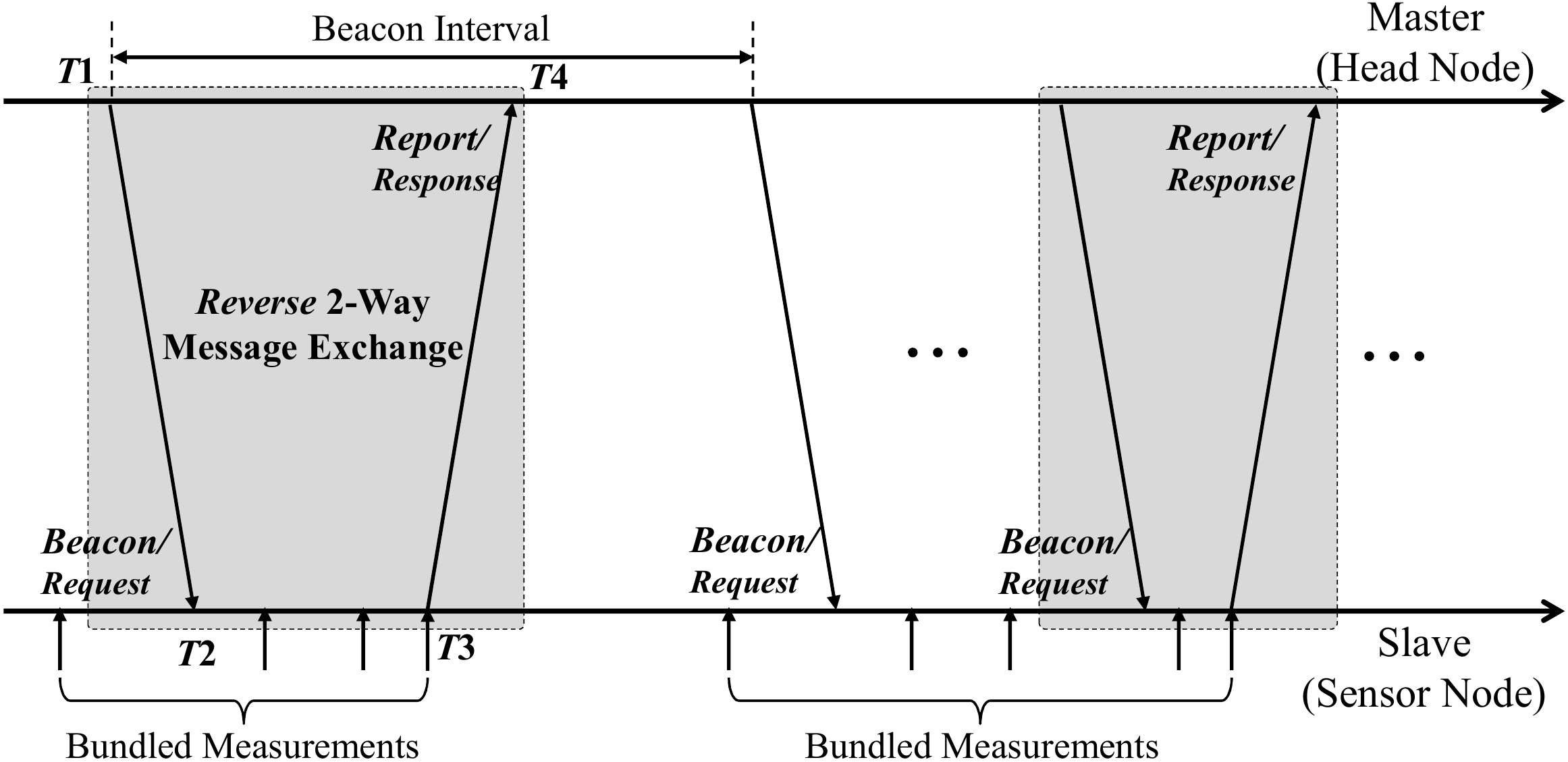}\\
    {\scriptsize (b)}
  \end{center}
  \caption{Comparison of two-way message exchanges: (a) Conventional two-way
    message exchanges as in the time-sync protocol for sensor networks (TPSN)
    \cite{ganeriwal03:_timin_protoc_sensor_networ}; (b) reverse two-way message
    exchanges of the proposed scheme shown with optional bundling of
    measurements in a ``Report/Response'' message.}
  \label{fig:proposed_scheme_overview}
\end{figure}

The major idea of the proposed scheme is to allow independent, unsynchronized
slave clocks at sensor nodes but running at the same frequency as the reference
clock at a head node through the asynchronous SCFR described in \cite{Kim:13-1},
which needs only the reception of messages with timestamps at sensor nodes. The
clock offset, on the other hand, is estimated at the head node based on the
reverse two-way message exchanges. Fig.~\ref{fig:proposed_scheme_overview}
illustrates this idea in comparison to conventional two-way message exchanges.
\begin{figure}[!tb]
  \centering
  \includegraphics[angle=-90,width=\linewidth]{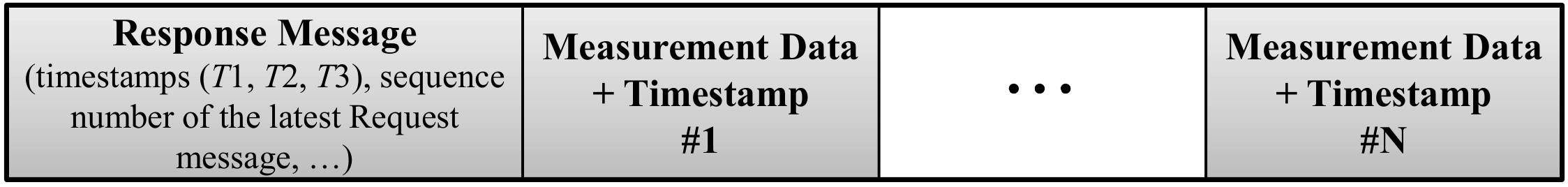}
  \caption{Payload structure of a "Report/Response" message of the proposed
    scheme with optional bundling of measurement data and timestamps.}
  \label{fig:payload_structure}
\end{figure}

First, the proposed scheme shown in Fig.~\ref{fig:proposed_scheme_overview}~(b)
does not have periodic, dedicated two-way message exchanges with synchronization
messages like ``Request'' and ``Response'' shown in
Fig.~\ref{fig:proposed_scheme_overview}~(a); instead, the ``Request'' and
``Response'' messages are embedded in the most recent timestamped
message---either broadcasted or unicasted to a specific node---from the head
node and a measurement data report message from a sensor node,
respectively. When there are no strict timing requirements for the processing of
measurement data, the measurement data and their corresponding timestamps can be
optionally bundled together in a ``Report/Response'' message, whose payload
structure is shown in Fig.~\ref{fig:payload_structure}, in order to further
reduce the number of message transmissions; in this case the clock offset
estimated based on the timestamp of the last measurement data \#N is
collectively applied to all bundled measurement data in estimating their
occurrences with respect to the reference clock at the head node.

Secondly, the direction of two-way message exchanges of the proposed scheme is
reversed, i.e., it is the master (i.e., the head node) that sends the
``Request'' messages, not the slave (i.e., the sensor node), unlike the
conventional two-way messages exchanges; as a result, the master knows the
current status of the slave clock, but the slave does not. So the information of
slave clocks (i.e., clock offsets with respect to the reference clock) is
centrally managed at the head node.

For operations like coordinated sleep wake-up node scheduling, before sending a
control message, the head node first adjusts the time for future operation based
on the clock offset of the recipient sensor node. In this way, even though
sensor nodes in the network have clocks with different clock offsets, their
operations can be coordinated based on the common reference clock at the head
node.

\subsection{Hardware and Logical Clock Models}
\label{usbsec:clock-models}
We consider an asymmetric WSN where one head (master) node and $N$ sensor
(slave) nodes all equipped with independent hardware clocks based on quartz
crystal oscillators. For simplicity, we take time $t$ of the head node clock as
a global reference and describe times of slave hardware clocks as functions of
$t$. We use the first-order affine clock model \cite{rajan11:_joint} to describe
time $T_{i}$ of the hardware clock at the $i$th sensor node as follows: For
$i \in [0,1,\ldots,N{-}1]$,
\begin{equation}
  \label{eq:hardware_clock_model}
  T_{i}(t) = (1 + \epsilon_{i})t + \theta_{i} ,
\end{equation}
where $(1{+}\epsilon_{i}){\in}\mathbb{R}_{+}$ and $\theta_{i}{\in}\mathbb{R}$
are clock frequency ratio and clock offset, respectively. Note that
$\epsilon_{i}$ is called a clock skew in the literature, which is defined as a
\textit{normalized clock frequency difference} between hardware clocks, and its
typical value for clocks based on quartz crystal oscillators is of the order of
tens of ppm (i.e., $\epsilon_{i}{\ll}1$).

WSN operations at a sensor node is based on a logical clock, whose time is again
a function of the time of its physical clock and takes into account the
adjustments by an adopted time synchronization scheme (e.g., offset adjustment
by two-way message exchanges and frequency adjustment by SCFR). Specifically,
time $\mathcal{T}_{i}$ of the logical clock at the $i$th sensor node can be
modeled as a piecewise linear function as follows\footnote{This model fits for
  WSN operations requiring times of discrete events only. More complicated
  (often nonlinear) models, however, is better suited for applications like
  playback of multimedia streaming where an analog or digital phase-locked loop
  (PLL) is used to generate a clock signal \cite{Kim:13-1}.}: For
$t_{k}{<}t{\leq}t_{k+1}$ ($k{=}0,1,\ldots$),
\begin{equation}
  \label{eq:logical_clock_model}
  \mathcal{T}_{i}\Big(T_{i}(t)\Big) = \mathcal{T}_{i}\Big(T_{i}(t_{k})\Big)
  + \dfrac{T_{i}(t)-T_{i}(t_{k})}{1 + \hat{\epsilon}_{i,k}} - \hat{\theta}_{i,k} ,
\end{equation}
where $t_{k}$ is the reference time when a $k$th synchronization occurs, and
$\hat{\epsilon}_{i,k}$ and $\hat{\theta}_{i,k}$ are the estimated clock skew and
offset from the $k$th synchronization. If the synchronization is only for
frequency, we set $\hat{\theta}_{i,k}$ to 0 in \eqref{eq:logical_clock_model};
if the synchronization is only for offset, on the other hand, we set
$\hat{\epsilon}_{i,k}$ to 0 in \eqref{eq:logical_clock_model}. It should be
noted that $\mathcal{T}_{i}$ in \eqref{eq:logical_clock_model} is the function
of $T_{i}(t)$, but not $t$ itself, because the reference time $t$ is not known
at a sensor node and the only available time is from a local hardware clock
(i.e., $T_{i}(t)$).


\subsection{Effect of Clock Skew on Measurement Time Estimation}
\label{sec:effect-clock-skew}
Here we compare the effect of clock skew on the measurement time estimation in
both conventional and reverse two-way message exchange procedures by an
approximate analysis of the best-case performance under a deterministic delay
(i.e., no random component) and no bundled measurements.
Fig.~\ref{fig:timing_diagrams} shows timing diagrams for the analysis of the
measurement time estimation error, where we assume the same amount of time
difference $T_{m}$ between the measurement occurrence and the reception of the
last time synchronization message---i.e., the ''Response'' message for the
conventional two-way message exchanges and a beacon (``Request'') message for
the reverse two-way message exchanges.
\setlength{\mylength}{.85\linewidth}
\begin{figure}[!tb]
  \begin{center}
    \includegraphics[angle=-90,width=\mylength]{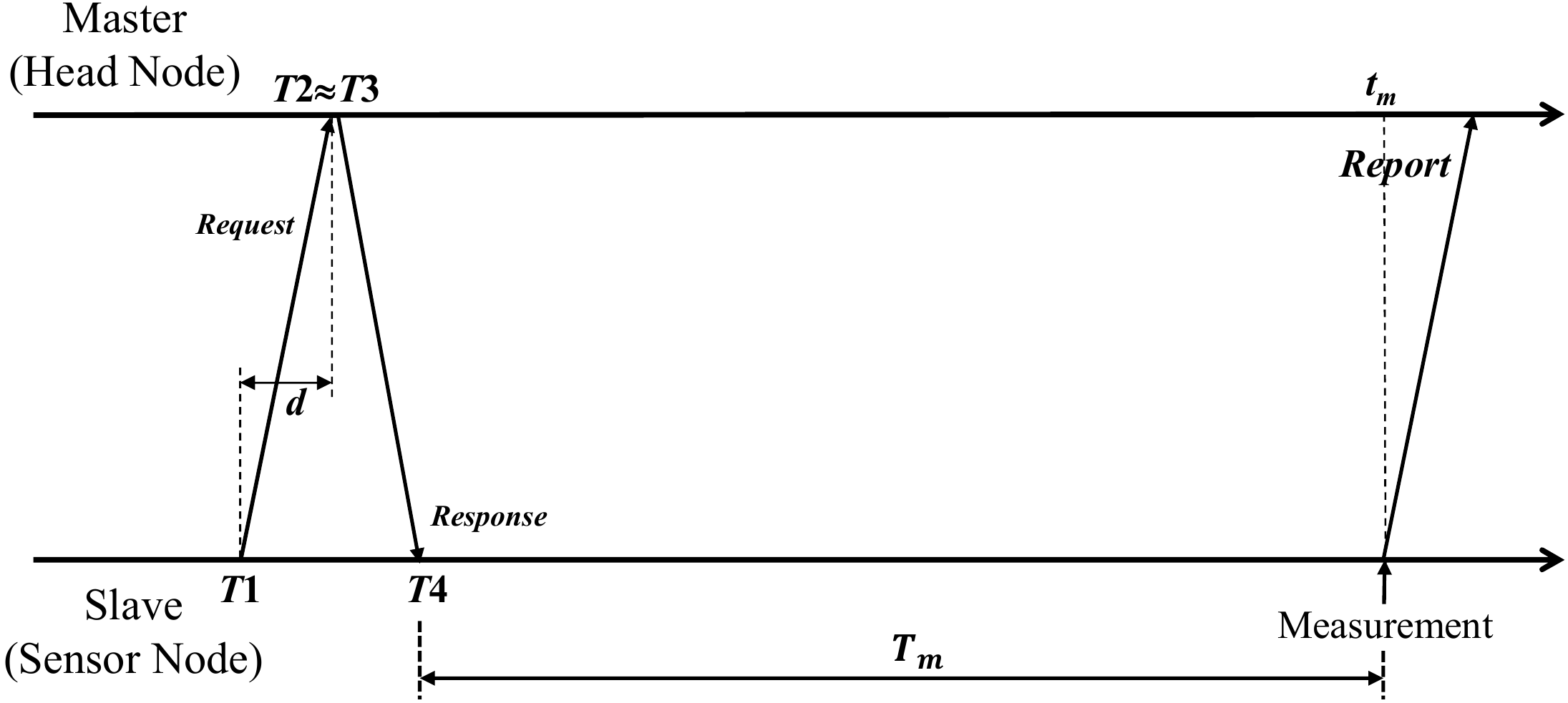}\\
    {\scriptsize (a)}\\
    \includegraphics[angle=-90,width=\mylength]{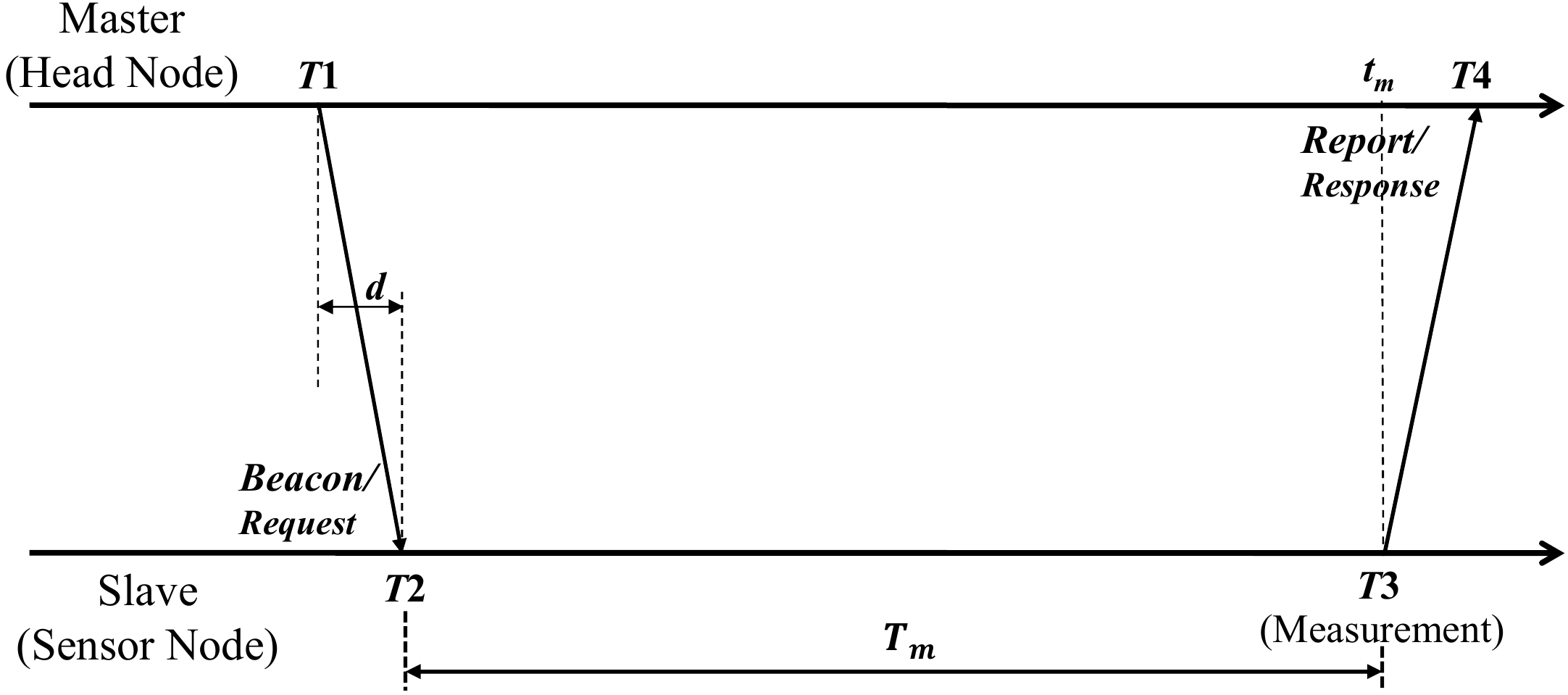}\\
    {\scriptsize (b)}
  \end{center}
  \caption{Timing diagrams for the analysis of the measurement time estimation
    error: (a) Conventional two-way message exchanges; (b) reverse two-way
    message exchanges of the proposed scheme.}
  \label{fig:timing_diagrams}
\end{figure}

Because MAC-layer timestamping of messages can remove all the sources of
uncertainties except the propagation delay \cite{akhlaq13:_rtsp} and the
variation in propagation delays is negligible in single-hop transmissions, the
clock offset estimation error $\Delta\hat{\theta}_{i}$ in the two-way message
exchanges without clock skew compensation (i.e., based on the hardware clock
model in \eqref{eq:hardware_clock_model}) is given by
\cite{ganeriwal03:_timin_protoc_sensor_networ}
\begin{equation}
  \Delta\hat{\theta}_{i} = \dfrac{(T4 - T1)\epsilon_{i}}{2} = d \times \epsilon_{i} ,
\end{equation}
where $d$ denotes one-way propagation delay. The measurement time estimation
error for the conventional two-way message exchanges shown in
Fig.~\ref{fig:timing_diagrams}~(a), therefore, can be expressed as
\begin{equation}
  \Delta\hat{t}_{m}^{Conv.} = d \epsilon_{i} + T_{m} \epsilon_{i} = (d + T_{m})\epsilon_{i} .
\end{equation}
If $T_{m}{\gg}d$, the measurement time error can be approximated as
\begin{equation}
  \label{eq:measurement_time_error_existing}
  \Delta\hat{t}_{m}^{Conv.} \sim T_{m} \times \epsilon_{i} .
\end{equation}
When the clock skew is compensated, \eqref{eq:measurement_time_error_existing}
becomes
\begin{equation}
  \label{eq:measurement_time_error_existing_skew_compensated}
  \Delta\hat{t}_{m}^{Conv.} \sim T_{m} \times \Delta\hat{\epsilon}_{i} ,
\end{equation}
where $\Delta\hat{\epsilon}_{i}$ is the clock skew estimation error.

For the reverse two-way message exchanges shown in
Fig.~\ref{fig:timing_diagrams}~(b), because the ``Request'' message of the
reverse two-way message exchanges is embedded in the measurement data ``Report''
message from a sensor node as shown in Fig.~\ref{fig:payload_structure}, there
is no time difference between the measurement time and the end of two-way
message exchange (i.e., $T_{4}$). In this case the two-way message exchange
procedure is the only source of error, i.e.,
\begin{equation}
  \Delta\hat{t}_{m}^{Rev.} = \dfrac{(T4 - T1)\Delta\hat{\epsilon}_{i}}{2} = \dfrac{(2d + T_{m})\Delta\hat{\epsilon}_{i}}{2} .
\end{equation}
Again, if $T_{m}{\gg}d$, the measurement time estimation error can be
approximated as
\begin{equation}
  \label{eq:measurement_time_error_proposed}
  \Delta\hat{t}_{m}^{Rev.} \sim \dfrac{T_{m}}{2} \times \Delta\hat{\epsilon}_{i} .
\end{equation}

From \eqref{eq:measurement_time_error_existing_skew_compensated} and
\eqref{eq:measurement_time_error_proposed}, we can see that, when being isolated
from the effect of random component of delay, the proposed scheme can reduce the
effect of clock skew on the measurement time estimation error by a factor of two
for $T_{m}{\gg}d$. Because the time difference $T_{m}$ between the reception of
the last beacon (or any timestamped message) from the head node and the
measurement occurrence at the sensor node can be a larger value in practice,
however, it is still important to compensate the clock skew at sensor nodes in
the proposed scheme.

\subsection{Asynchronous SCFR at Sensor Nodes: One-Way Clock Skew Estimation}
\label{sec:async-scfr}
One of the essential components of the proposed time synchronization scheme is
the recovery of the reference clock frequency at sensor nodes based on one-way
message dissemination from the head node. Once we estimate the clock skew
$\epsilon_{i}$ in \eqref{eq:hardware_clock_model}, the reference clock frequency
can be recovered through the clock skew compensation of the logical clock model
in \eqref{eq:logical_clock_model}.

In case of two-way message exchanges, both joint ML estimation of clock offset
and skew with a known fixed portion of delay and separate ML-like estimation of
clock skew are studied in \cite{noh07:_novel}. As for one-way message
dissemination, here we derive both joint ML and separate ML-like estimators in a
similar manner but based on the problem formulation in \cite{Kim:13-1}.

Let $t_{d}(k)$ ($k{=}0,1,\ldots$) be the reference time for the $k$th message's
departure from the head node; note that $t_{d}(k)$ also denotes the value of the
timestamp carried by the $k$th message. From the hardware clock model in
\eqref{eq:hardware_clock_model}, we can obtain $t_{a,i}(k)$, the arrival time of
the $k$th message with respect to the $i$th sensor node's \textit{hardware
  clock}, as follows:
\begin{align}
  \label{eq:arrival_timestamp}
  t_{a,i}(k) & = T_{i}\left(t_{d}(k)\right) + d(k) \notag \\
             & = (1 + \epsilon_{i})t_{d}(k) + \theta_{i} + d(k) ,
\end{align}
where $d(k)$ denotes a one-way packet delay from the head node to the $i$th
sensor node in terms of the $i$th sensor node's hardware clock.\footnote{This
  definition of the one-way delay (i.e., in terms of the sensor nodes' hardware
  clock) is different from that in \cite{noh07:_novel} and
  \cite{guchhait15:_joint_minim_varian_unbias_maxim}, where the delay is defined
  in terms of the head node clock, and makes simpler the derivation of the
  one-way estimators in this paper.} For the observation of timestamps from
\eqref{eq:arrival_timestamp}, we can obtain the joint ML estimators (MLEs) of
clock offset and skew as stated in Proposition~\ref{th:joint_mle}.
\begin{theorem}
  \label{th:joint_mle}
  For a white Gaussian delay $d(k)$ with \textit{known} mean $d$ and variance
  $\sigma^{2}$, the joint ML estimators (MLEs) for clock offset
  $\hat{\theta}^{ML}_{i}(k)$ and skew $\hat{R}^{ML}_{i}(k)$ in
  \eqref{eq:arrival_timestamp} are given by
  \begin{equation}
    \label{eq:joint_mle_offset_closed_form}
    \hat{\theta}^{ML}_{i}(k) = \dfrac{
      \overline{t^{2}_{d}} \cdot \overline{t_{a,i}} - \overline{t_{d}} \cdot \overline{t_{d}t_{a,i}}
    }{
      \overline{t^{2}_{d}} - \left(\overline{t_{d}}\right)^{2}
    } - d ,
  \end{equation}
  \begin{equation}
    \label{eq:joint_mle_skew_closed_form}
    \hat{R}^{ML}_{i}(k) = \dfrac{
      \overline{t_{d}t_{a,i}} - \overline{t_{d}} \cdot \overline{t_{a,i}}
    }{
      \overline{t^{2}_{d}} - \left(\overline{t_{d}} \right)^{2}} ,
  \end{equation}
  where the notations $\overline{x}$ and $\overline{xy}$ denote the average
  values of $x(j)$ (i.e., $\sum^{k}_{j=0}x(j)/k$) and $x(j)y(j)$ (i.e.,
  $\sum^{k}_{j=0}x(j)y(j)/k$), respectively. Also $\hat{\theta}^{ML}_{i}(k)$ and
  $\hat{R}^{ML}_{i}(k)$ are \textit{efficient estimators}
  \cite[p.~34]{kay93:_fundam} which are unbiased and attain the Cram\'{e}r-Rao
  lower bounds (CRLBs) given by
  \begin{equation}
    \label{eq:joint_mle_offset_crlb}
    \operatorname{Var}\left(\hat{\theta}_{i}(k)\right) \geq
    \dfrac{
      \sigma^{2} \cdot \overline{t_{d}^{2}}
    }{
      k\left\{\overline{t_{d}^{2}}-\left(\overline{t_{d}}\right)^{2}\right\}
    } ,
  \end{equation}
  \begin{equation}
    \label{eq:joint_mle_skew_crlb}
    \operatorname{Var}\left(\hat{R}_{i}(k)\right) \geq
    \dfrac{
      \sigma^{2}
    }{
      k\left\{\overline{t_{d}^{2}}-\left(\overline{t_{d}}\right)^{2}\right\}
    } .
  \end{equation}
\end{theorem}
\begin{IEEEproof}
  The proof of Proposition~\ref{th:joint_mle} is presented in
  Appendix~\ref{sec:proof-proposition-1}.
\end{IEEEproof}
Note that, even though the mean and the variance of white Gaussian delay are
assumed to be known for the derivation of the joint MLEs, the resulting clock
skew estimator $\hat{R}^{ML}_{i}(k)$---the only one needed in the proposed
scheme---does not depend on it.

Compared to the joint MLEs for two-way message exchanges derived in
\cite{noh07:_novel}, \eqref{eq:joint_mle_offset_closed_form} and
\eqref{eq:joint_mle_skew_closed_form} take simpler expressions, but they are
still complicated and not suitable for recursive implementation.
Because we are only interested in the estimation of the clock
skew---equivalently the ratio of clock frequencies---at sensor nodes, we can
formulate the estimation problem where the clock skew is the only parameter to
estimate. In \cite{Kim:13-1}, the problem of asynchronous SCFR is formulated as
a linear \textit{regression through the origin (RTO)} model by subtracting both
sides of \eqref{eq:arrival_timestamp} with their initial values: For
$k=1,2,\ldots$,
\begin{equation}
  \label{eq:rot_model}
  \tilde{t}_{a,i}(k) = (1 + \epsilon_{i})\tilde{t}_{d}(k) + \tilde{d}(k) ,
\end{equation}
where $\tilde{t}_{a,i}(k){\triangleq}t_{a,i}(k){-}t_{a,i}(0)$,
$\tilde{t}_{d}(k){\triangleq}t_{d}(k){-}t_{d}(0)$, and
$\tilde{d}(k){\triangleq}d(k){-}d(0)$. Note that $\tilde{d}(k)$ now represents a
noise process with a zero mean for a stationary delay model.

Note that $\tilde{t}_{a,i}(k)$'s in \eqref{eq:rot_model} are not independent one
another due to $\tilde{d}(k)$. The derivation of an MLE based on all
observations of $\tilde{t}_{a,i}(k)$'s, therefore, is not straightforward. In
\cite{Kim:13-1}, two practical estimators are proposed in this regard, i.e., the
recursive least squares (RLS) and the cumulative ratio (CR), which are not based
on any assumption on the delay distribution. Of the two estimators, the CR
estimator is best suited for battery-powered sensor nodes in the proposed time
synchronization scheme due to its lower complexity. If we define $R_{i}$ as the
ratio of the $i$th sensor node hardware clock frequency to that of the reference
clock (i.e., $1{+}\epsilon_{i}$), the CR estimator $\hat{R}^{CR}_{i}(k)$ is
given by
\begin{equation}
  \label{eq:cr_iter}
  \hat{R}^{CR}_{i}(k) = \dfrac{\tilde{t}_{a,i}(k)}{\tilde{t}_{d}(k)} .
\end{equation}
From \eqref{eq:rot_model}, we can see that $\hat{R}^{CR}_{i}(k)$ can be
rewritten as follows:
\begin{equation}
  \label{eq:cr}
  \hat{R}^{CR}_{i}(k) = R_{i} + \dfrac{\tilde{d}(k)}{\tilde{t}_s(k)} ,
\end{equation}
where its noise component becomes zero as time goes to infinity irrespective of
its statistical characteristics. Unlike the RLS estimator, there are no design
parameter values or initial values to set for the CR estimator.

Note that for a Gaussian delay model, the CR estimator becomes an unbiased
estimator achieving its lower bound as stated in Proposition~\ref{th:ee_skew}.
\begin{theorem}
  \label{th:ee_skew}
  For a white Gaussian delay $\tilde{d}(k)$ with zero mean and variance
  $\sigma^{2}$, the CR estimator in \eqref{eq:cr} becomes an \textit{unbiased
    estimator} which attains the lower bound\footnote{This lower bound is in
    fact the CRLB for the case when the observations are limited to the
    timestamps from the first and last messages.} given by
  \[
  \operatorname{Var}\left(\hat{R}^{CR}_{i}(k)\right) \geq
  \dfrac{2\sigma^{2}}{\tilde{t}_{d}(k)^{2}} .
  \]
\end{theorem}

\begin{IEEEproof}
  The proof of Proposition~\ref{th:ee_skew} is presented in
  Appendix~\ref{sec:proof-proposition-2}.
\end{IEEEproof}

\subsection{Extension to Multi-Hop Time Synchronization}
\label{sec:extension-multi-hop}
Fig.~\ref{fig:multi-hop_extension} shows how the proposed scheme can be extended
to a hierarchical structure for network-wide, multi-hop synchronization through
simple packet-relaying or more advanced time-translating gateway nodes, the
latter of which act as both head nodes (for the nodes in the lower hierarchy)
and normal sensor nodes (for the gateway node in the higher hierarchy):
\setlength{\mylength}{.65\linewidth}
\begin{figure}[!tb]
  \begin{center}
    \includegraphics[width=\mylength]{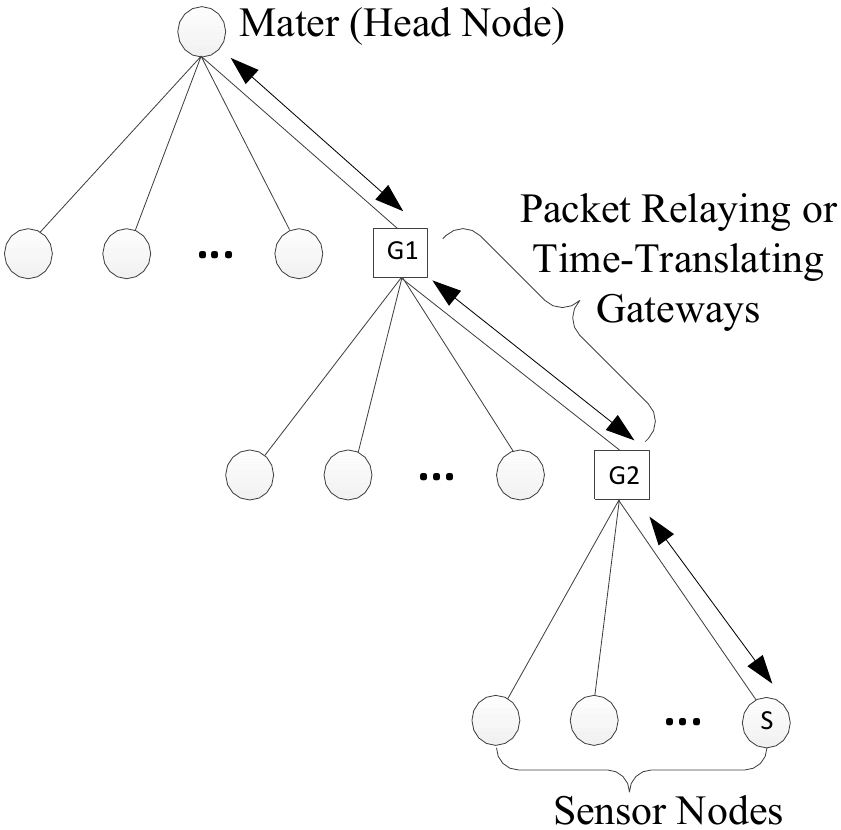}
  \end{center}
  \caption{Extension of the proposed time synchronization scheme to a
    hierarchical structure for network-wide, multi-hop synchronization through
    packet-relaying or time-translating gateway nodes.}
  \label{fig:multi-hop_extension}
\end{figure}
For example, consider the message transmission from the sensor node S to the
head node through the two gateway nodes G1 and G2 as shown in
Fig.~\ref{fig:multi-hop_extension}. In case of \textit{packet relay}, G2 simply
passes the received timestamped message from S to G1 and G1 to the head node
without any change of the value of timestamp in the middle. In case of
\textit{time translation}, because G2 acts as a head node for S, it first
translates the value of timestamp based on the information on the clock offset
of S. Then G2 relays the message from S to G1 with translated timestamp value
(with respect to its own clock). From G1's point of view, G2 is just one of
sensor nodes it manages. Again, based on the information on the clock offset of
G2, G1 translates the value of clock offset with respect to its own clock and
relays the message to the head node. Finally, the head node receives the message
from S, which is just relayed by G1, and translates the timestamp value based on
the information on the clock offset of G1 it manages. In this way the head node
can obtain the measurement data and its occurrence in time reported by S with
respect to its own reference clock.

It is clear that, compared to time-translating gateway nodes, packet-relaying
gateway nodes could be much simpler because there is no additional functionality
(i.e., estimation and management of the offset information and translation of
timestamp values) except packet relaying. A downside, however, is that they
introduce larger packet delays resulting from queueing and MAC operations, which
could deteriorate the performance of time synchronization.

\section{Simulation Results}
\label{sec:simulation-results}
We carry out a series of simulation experiments to investigate the performance
of the proposed time synchronization scheme in terms of the accuracy of
measurement time estimation and the number of message transmissions and
receptions at a sensor node in comparison with those based on the conventional
two-way message exchanges. We also analyze the performance of one-way clock skew
estimators independently of the proposed time synchronization scheme to choose
the best one for battery-powered sensor nodes.

For the simulations, we consider a simple WSN with one head node and one sensor
node that are deployed \SI{100}{\meter} from each other because the time
synchronization of a sensor node in the proposed scheme can be carried out
independently of that of other sensor nodes. As in \cite{chepuri13:_joint}, we
set the clock frequency ratio $R_{i}$ and the clock offset $\theta_{i}$ to 1 +
100 ppm and \SI{1}{\second}, respectively. We model propagation delay, which
takes into account timestamp generation and reception noise as well, with an
independent and identically distributed (i.i.d.) Gaussian process unless
specified otherwise.

\subsection{Performance of One-Way Clock Skew Estimation}
\label{sec:performance-one-way}
First, we analyze the performance of one-way clock skew estimators discussed in
Section~\ref{sec:async-scfr}, together with the two-way ML-like estimator for
Gaussian delay model (GMLLE) proposed in \cite{noh07:_novel} for comparison.
Fig.~\ref{fig:skew_est_normal} shows the mean square error (MSE) of both one-way
and two-way clock skew estimators with Gaussian delays with standard deviation
of \SI{1}{\nano\second} and \SI{1}{\micro\second}, which are calculated over
10,000 simulation runs. Fig.~\ref{fig:skew_est_ar1} also shows the results with
first-order autoregressive (AR(1)) random delays with correlation coefficient
($\rho$) of 0.6 and standard deviation of \SI{1}{\micro\second} and
\SI{1}{\milli\second}; the results in this case demonstrate the performance of
the estimators under correlated delays, which reflect the case of multi-hop
extension through packet-relaying gateway nodes as discussed in
Section~\ref{sec:extension-multi-hop}.
\setlength{\mylength}{.85\linewidth} \setlength{\myvspace}{-0.3cm}
\begin{figure*}[!tb]
  \begin{minipage}{.49\linewidth}
    \begin{center}
      \includegraphics[width=\mylength]{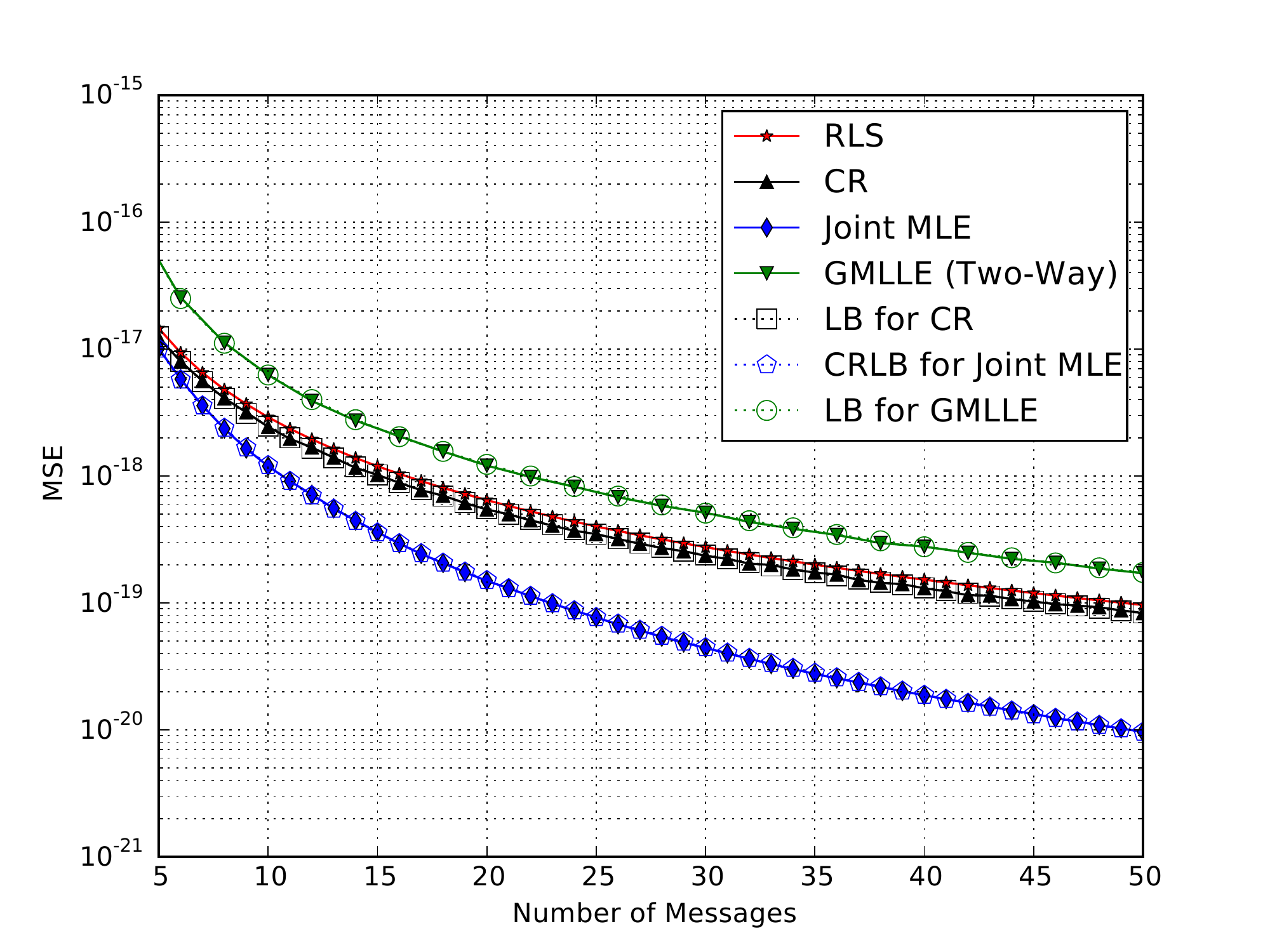}\\
      \vspace{\myvspace}%
      {\scriptsize (a)}
    \end{center}
  \end{minipage}
  \hfill
  \begin{minipage}{.49\linewidth}
    \begin{center}
      \includegraphics[width=\mylength]{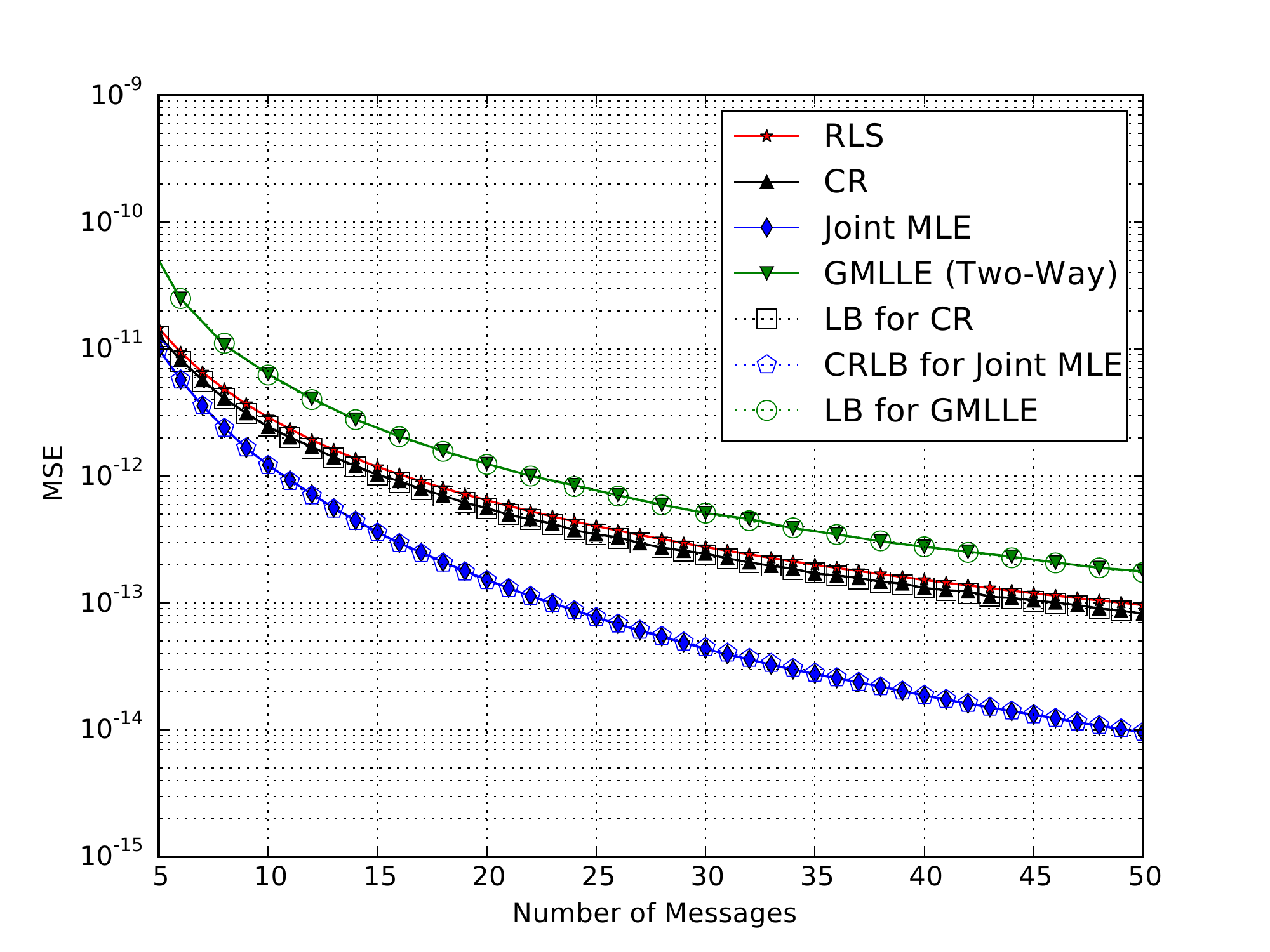}\\
      \vspace{\myvspace}%
      {\scriptsize (b)}
    \end{center}
  \end{minipage}
  \caption{MSE of the estimated clock skews from one-way and two-way clock skew
    estimators for Gaussian random delays with standard deviation ($\sigma$) of
    (a) \SI{1}{\nano\second} and (b) \SI{1}{\micro\second}.}
  \label{fig:skew_est_normal}
\end{figure*}
\begin{figure*}[!tb]
  \begin{minipage}{.49\linewidth}
    \begin{center}
      \includegraphics[width=\mylength]{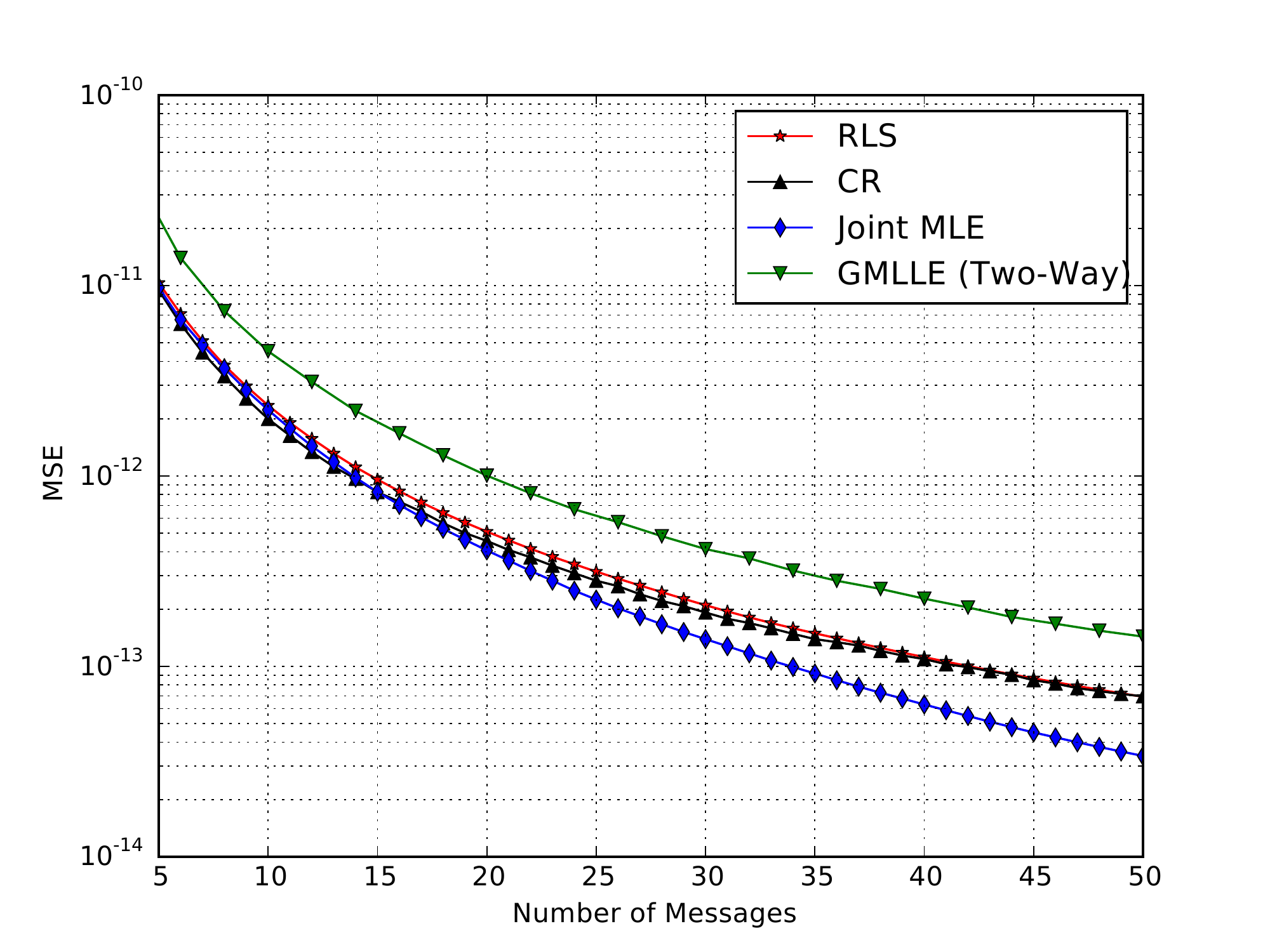}\\
      \vspace{\myvspace}%
      {\scriptsize (a)}
    \end{center}
  \end{minipage}
  \hfill
  \begin{minipage}{.49\linewidth}
    \begin{center}
      \includegraphics[width=\mylength]{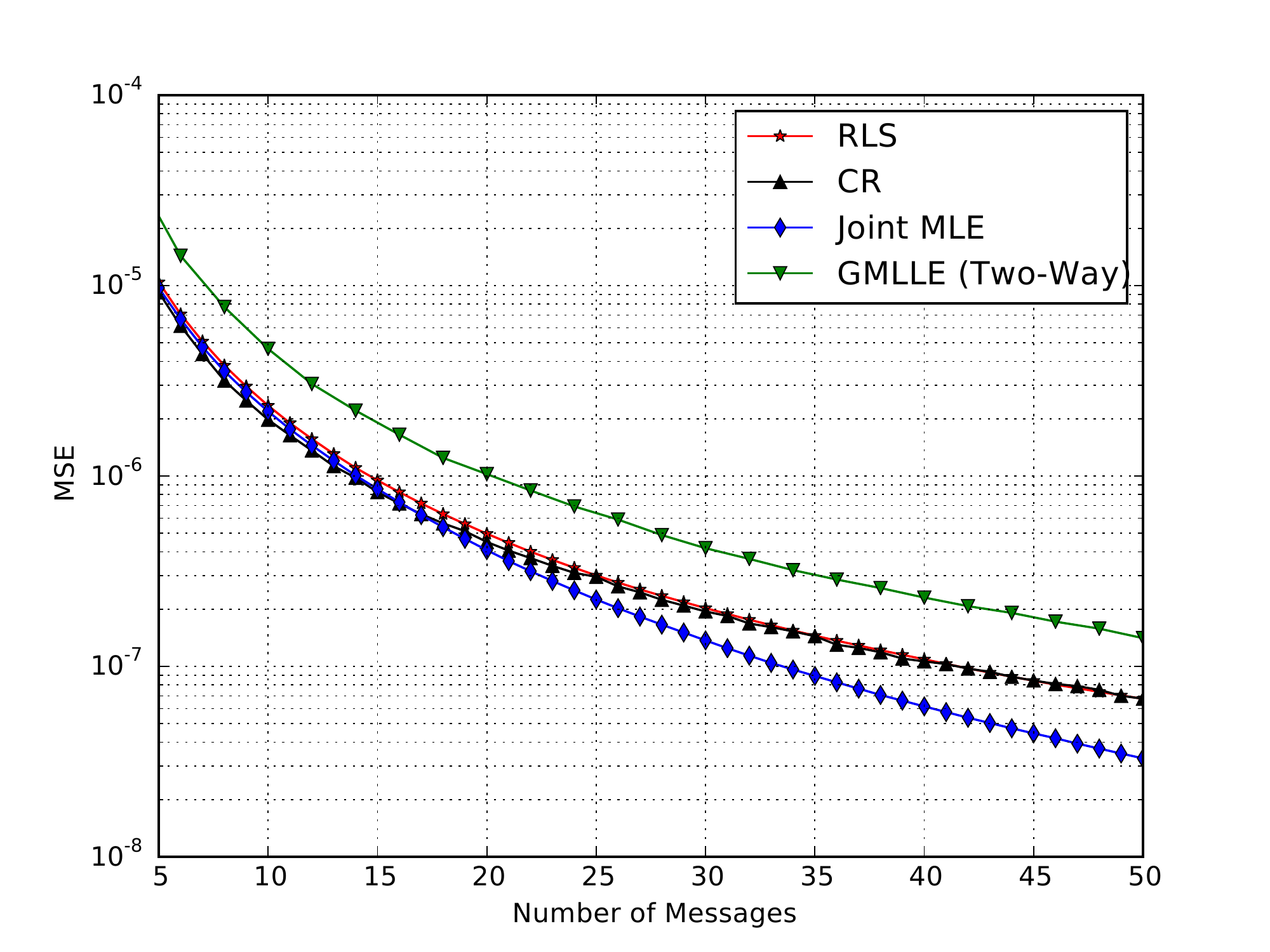}\\
      \vspace{\myvspace}%
      {\scriptsize (b)}
    \end{center}
  \end{minipage}
  \caption{MSE of the estimated clock skews from one-way and two-way clock skew
    estimators for AR(1) random delays with correlation coefficient ($\rho$) of
    0.6 and standard deviation of (a) \SI{1}{\micro\second} and (b)
    \SI{1}{\milli\second}.}
  \label{fig:skew_est_ar1}
\end{figure*}


The results for both Gaussian and AR(1) delays show that the joint MLE provides
the best performance as expected because it can use all observed timestamp
values. The performance gap between the joint MLE and other one-way estimators
(i.e., CR and RLS), however, becomes narrower when the delay does not follow the
Gaussian distribution and has correlation. The estimation performance of the
two-way scheme (i.e., GMLLE), on the other hand, is the worst \textit{given the
  number of messages exchanged}, which indicates that the two-way message
exchange procedure is not energy-efficient in estimating the clock skew only.

As discussed in Section~\ref{sec:async-scfr}, due to its lower complexity and
robustness to delay characteristics, we use the CR estimator at battery-powered
sensor nodes in the investigation of the performance of the proposed time
synchronization scheme in the following sections.

\subsection{Performance of Measurement Time Estimation and Energy Efficiency}
\label{sec:meas-time-estim}
For the analysis of the performance of the proposed time synchronization scheme,
we run simulations for three different values of synchronization interval
(SI)\footnote{This is the interval of beacons for the proposed scheme and the
  interval of two-way message exchanges for others.}---i.e., \SI{100}{\second},
\SI{1}{\second} and \SI{10}{\milli\second}---to investigate the effect of the
time difference $T_{m}$ between the last time synchronization message from the
head node and the measurement occurrence at the sensor node. The propagation
delay is modeled as an i.i.d. Gaussian process with a standard deviation of
\SI{1}{\nano\second}. During the observation interval of \SI{1}{\hour}, total
100 measurements are made where their corresponding data arrivals are modeled as
a Poisson process.  For SCFR, we use the CR estimator for the proposed scheme
based on the observation in Section~\ref{sec:performance-one-way} and the GMLLE
for the two-way scheme for comparison.

The results of SCFR at the sensor node and measurement time estimation at the
head node (i.e., the estimation of the measurement time at the sensor node with
respect to the reference clock at the head node) are shown in
Fig.~\ref{fig:time_sync_results_si}, and their MSEs calculated over samples from
one simulation run with the number of message transmissions (N$_{\rm TX}$) and
receptions (N$_{\rm RX}$) at the sensor node are summarized in
Table~\ref{tab:time_sync_results_si}.
\setlength{\mylength}{.85\linewidth} \setlength{\myvspace}{-0.3cm}
\begin{figure}[!tb]
  \begin{center}
    \includegraphics[width=\mylength]{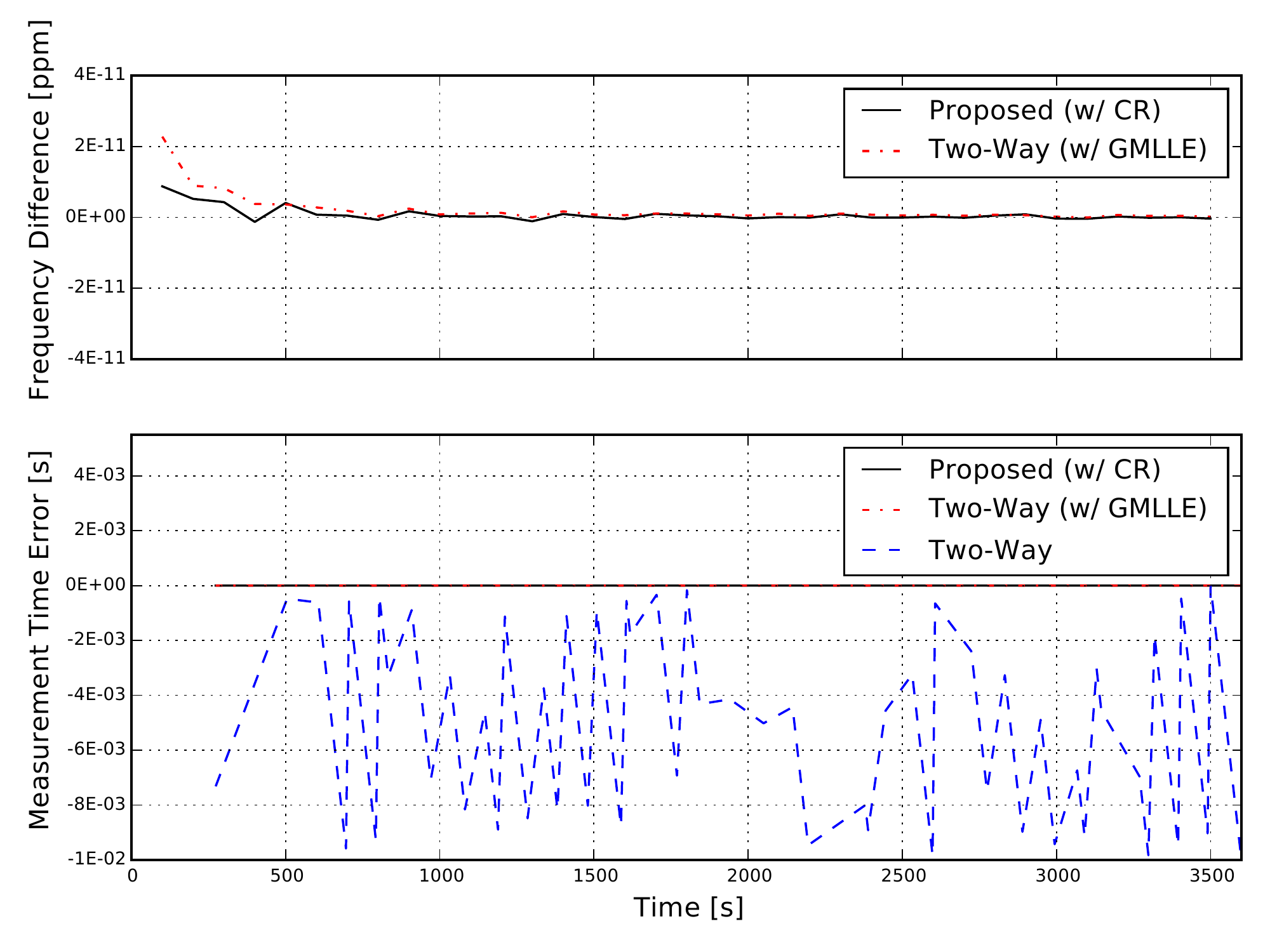}\\
    \vspace{\myvspace}
    {\scriptsize (a)}\\
    \includegraphics[width=\mylength]{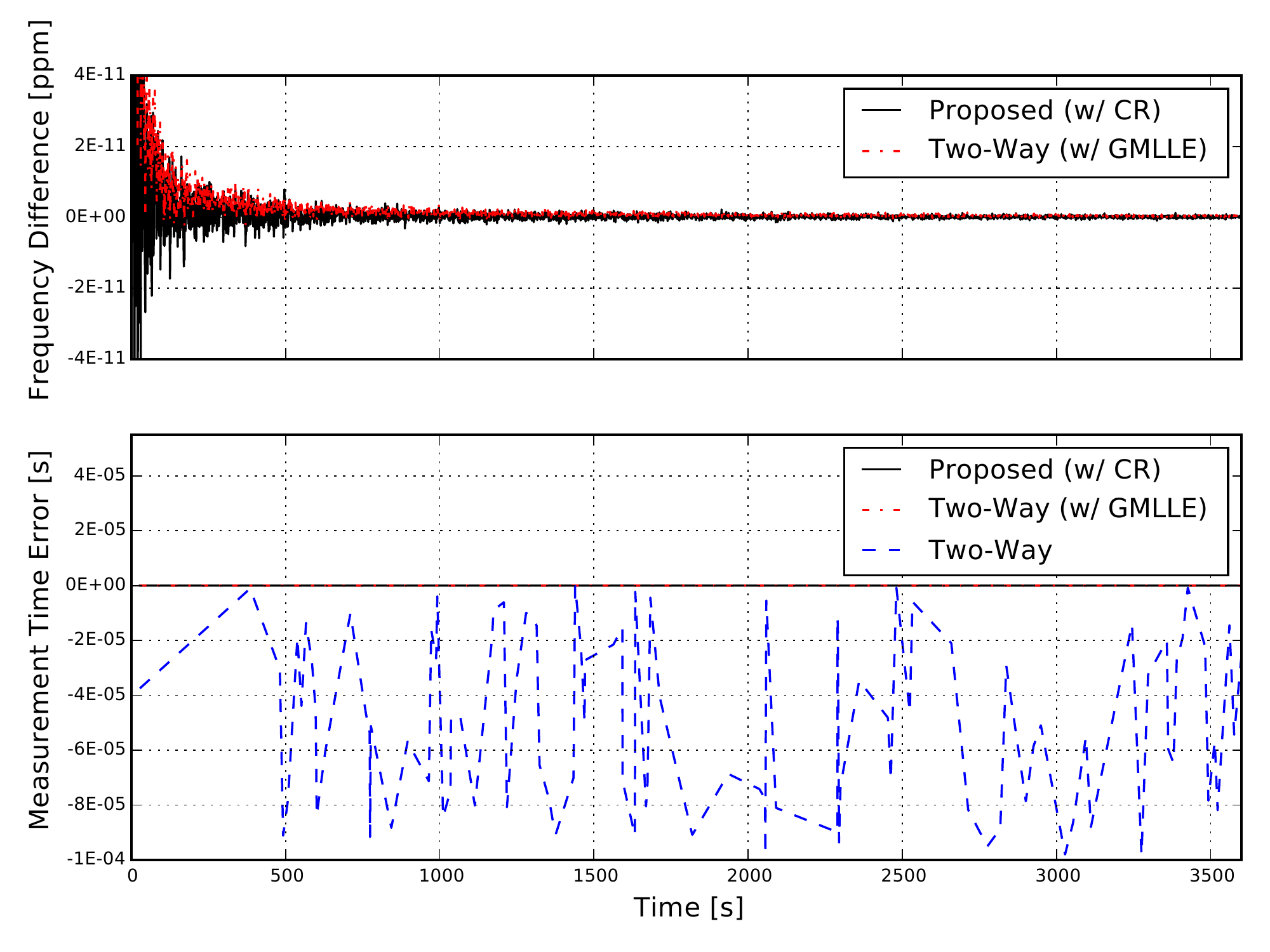}\\
    \vspace{\myvspace}
    {\scriptsize (b)}\\
    \includegraphics[width=\mylength]{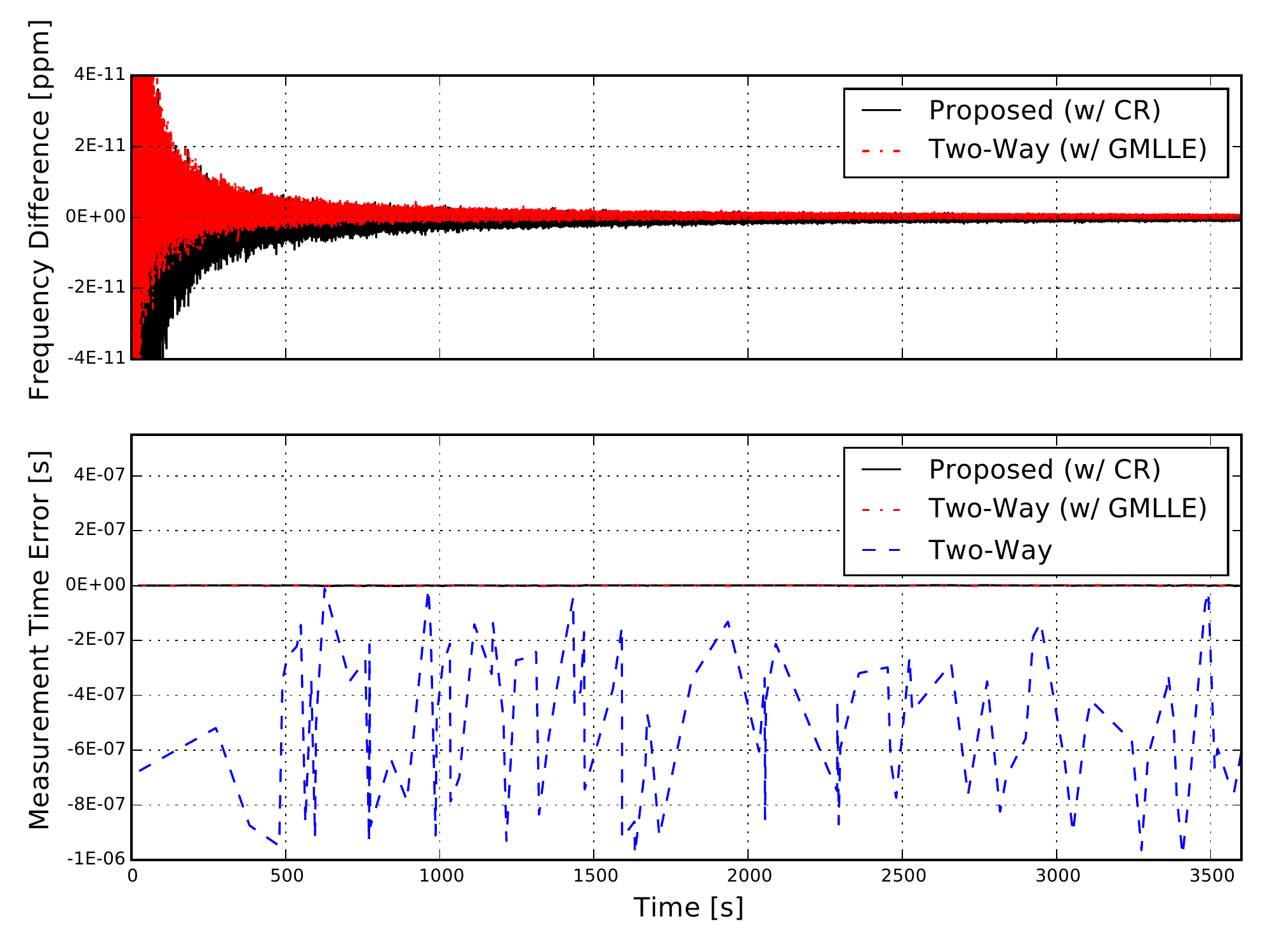}\\
    \vspace{\myvspace}%
    {\scriptsize (c)}
  \end{center}
  \caption{Performance of SCFR at a sensor node and measurement time estimation
    at a head node for SI of (a) \SI{100}{\second}, (b) \SI{1}{\second}, and (c)
    \SI{10}{\milli\second}.}
  \label{fig:time_sync_results_si}
\end{figure}
\begin{table*}[!tb]
  \centering
  \begin{threeparttable}
    \caption{Results of Time Synchronization Simulations for Different SIs}
    \label{tab:time_sync_results_si}
    \centering
    \begin{tabular}{|c|l||r|r|r|r|}
      \hline
      \multicolumn{2}{|c||}{Synchronization Scheme}
      & Skew Estimation MSE\tnote{1}
      & Measurement Time Estimation MSE\tnote{1}
      & N$_{\rm TX}$ & N$_{\rm RX}$ \\ \hline\hline
      \multirow{3}{*}{Proposed}
      & SI$\;=100~\mbox{s}$ & 8.8811E-25 & 5.8990E-19 & 100 & 36 \\ \cline{2-6}
      & SI$\;=1~\mbox{s}$ & 9.1748E-25 & 5.4210E-19 & 100 & 3600 \\ \cline{2-6}
      & SI$\;=10~\mbox{ms}$ & 1.0887E-24 & 4.7684E-19 & 100 & 360100 \\ \hline\hline
      \multirow{3}{*}{Two-Way with GMLLE}
      & SI$\;=100~\mbox{s}$ & 1.9021E-24 & 4.7784E-19 & 136 & 36 \\ \cline{2-6}
      & SI$\;=1~\mbox{s}$ & 1.7034E-24 & 6.1452E-19 & 3700 & 3600 \\ \cline{2-6}
      & SI$\;=10~\mbox{ms}$ & 9.0992E-25 & 4.0485E-19 & 360100 & 360000 \\
      \hline\hline
      \multirow{3}{*}{Two-Way}
      & SI$\;=100~\mbox{s}$ & \multicolumn{1}{c|}{\multirow{3}{*}{N/A}} & 3.4900E-05 & 136 & 36 \\ \cline{2-2}\cline{4-6}
      & SI$\;=1~\mbox{s}$ & & 3.4564E-09 & 3700 & 3600 \\ \cline{2-2}\cline{4-6}
      & SI$\;=10~\mbox{ms}$ & & 3.3638E-13 & 360100 & 360000 \\ \hline
    \end{tabular}
    \begin{tablenotes}
    \item[1] For the samples obtained from \SI{360}{\second} (i.e., one tenth of
      the observation period) to avoid the effect of a transient period.
    \end{tablenotes}
  \end{threeparttable}
\end{table*}

As shown by the approximate analysis in Section~\ref{sec:effect-clock-skew}, the
results show that, without SCFR (i.e., ``Two-Way''), the measurement time
estimation errors highly depend on the synchronization interval. The use of SCFR
(i.e., ``Proposed (w/ CR)'' and ``Two-Way (w/ GMLLE)''), on the other hand,
greatly reduces this dependency, and the resulting measurement time estimation
errors (i.e., the square root of MSE of measurement time estimation) are of the
order of the noise standard deviation (i.e., \SI{1}{\nano\second}) for all
values of the synchronization interval, which means that the synchronization
interval can be increased without much affecting the time synchronization
performance to save energy. Table~\ref{tab:time_sync_results_si} also shows that
the number of message transmissions for the proposed scheme is the same as the
number of measurements irrespective of the synchronization interval, while the
number of message receptions increases as the synchronization interval decreases
for all the schemes considered; in case of the synchronization interval of
\SI{1}{\second}, the number of message transmissions is 100 for the proposed
scheme and 3,700 (100 for the measurement data and 3,600 for the synchronization
messages) for the schemes based on the conventional two-way messages exchanges.


\subsection{Effect of Bundling of Measurement Data}
\label{sec:effect-bundl-meas}
\begin{figure}[!tb]
  \begin{center}
    \includegraphics[width=.85\linewidth]{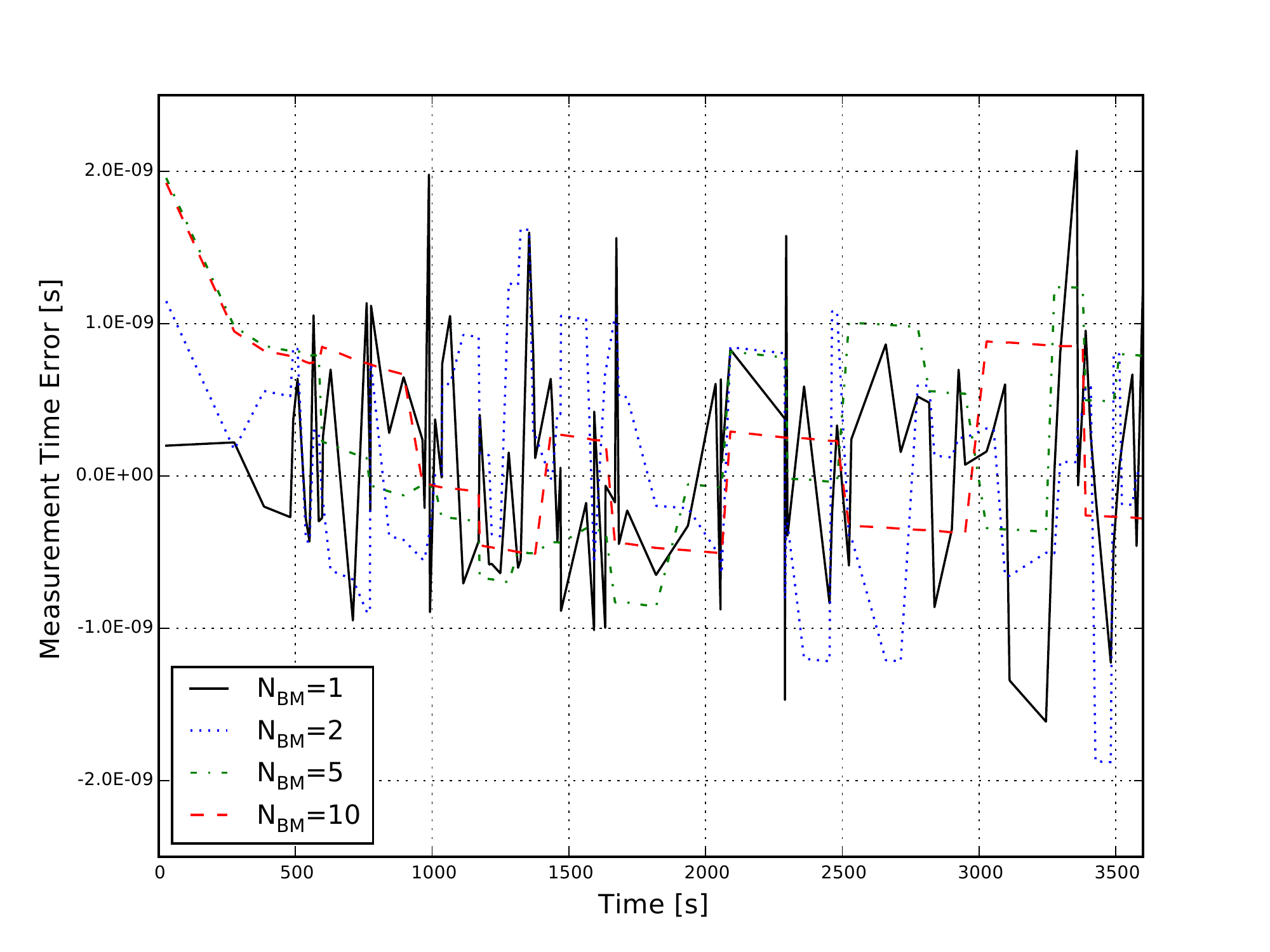}
  \end{center}
  \caption{Measurement time estimation errors for different numbers of bundled
    measurements (N$_{\rm BM}$) with SI of 1 s.}
  \label{fig:time_sync_results_nb}
\end{figure}

We also investigate the effect of the bundling of measurement data described in
Section~\ref{sec:energy-effic-time} with fixed synchronization interval of
\SI{1}{\second}. Fig.~\ref{fig:time_sync_results_nb} shows the measurement time
estimation errors for different numbers of bundled measurements (N$_{\rm BM}$)
in one ``Report/Response'' message, and again their MSEs calculated over samples
from one simulation run with the number of message transmissions and receptions
at the sensor node are summarized in Table~\ref{tab:time_sync_results_nb}.
\begin{table*}[!tb]
  \centering
  \begin{threeparttable}
    \caption{Results of Time Synchronization Simulations for Different Numbers
      of Bundled Measurements (N$_{\rm BM}$) with SI of 1 s.}
    \label{tab:time_sync_results_nb}
    \centering
    \begin{tabular}{|c|l||r|r|r|}
      \hline
      \multicolumn{2}{|c||}{Synchronization Scheme}
      & Measurement Time Estimation MSE\tnote{1}
      & N$_{\rm TX}$ & N$_{\rm RX}$ \\ \hline\hline
      \multirow{4}{*}{Proposed}
      & N$_{\rm BM}=1$ & 5.4210E-19 & 100 & 3600 \\ \cline{2-5}
      & N$_{\rm BM}=2$ & 5.1116E-19 & 50 & 3600 \\ \cline{2-5}
      & N$_{\rm BM}=5$ & 3.7504E-19 & 20 & 3600 \\ \cline{2-5}
      & N$_{\rm BM}=10$ & 2.6468E-19 & 10 & 3600 \\ \hline
    \end{tabular}
    \begin{tablenotes}
    \item[1] For the samples obtained from \SI{360}{\second} (i.e., one tenth of
      the observation period) to avoid the effect of a transient period.
    \end{tablenotes}
  \end{threeparttable}
\end{table*}

It is evident from the figure and the table that the measurement time estimation
errors are not much affected by the number of bundled measurements after the
initial transient period (i.e., after around \SI{500}{\second}). In fact, the
table shows that the MSE of measurement time estimation slightly decreases as
the number of bundled measurements increases; this is because all the
measurement data reported in a ``Report/Response'' message share the same
estimated value of the clock offset as the last measurement in the bundle, which
in fact reduces the fluctuations in the measurement time estimation errors as
observed in Fig.~\ref{fig:time_sync_results_nb}.

The bundling of measurement data further reduces the number of message
transmissions: For instance, if we bundle 10 measurement data in a
``Report/Response'' message (i.e., N$_{\rm BM}{=}10$), we can reduce the number
of message transmissions from 100 to 10, which is less than 0.3 percentage of
the number of message transmissions for the schemes based on the conventional
two-way message exchanges (i.e., 3,700).

Note that we need to be careful in interpreting the number of message
transmissions as an indirect measure of energy consumption in this case. Because
the bundling increases the length of message payload, the total number of bits
transmitted for the case with the bundling is not the same as that for the case
without the bundling given the number of messages. Still, however, the
measurement bundling is an effective option to save the energy consumption
because we can save the energy for the transmission of the overhead of a message
including a preamble, a frame header, and a frame check sequence by reducing the
number of message transmissions. Also note that the measurement bundling can be
used only when there are no strict timing requirements for the processing of
measurement data due to the increased delay resulting from the bundling process.

\section{Comparison to Related Work}
\label{sec:comp-relat-work}
\textit{Unsynchronized Clocks}: The proposed time synchronization scheme is
similar to the \textit{post-facto synchronization} scheme described in
\cite{elson01:_time} in that sensor nodes' clocks are unsynchronized and that
the time of the occurrence of an event is first recorded with respect to the
local clock. The difference is that time translation occurs at sensor nodes just
after receiving synchronization messages from a ``third-party'' node (e.g., a
head node) in the post-facto synchronization, while, in the proposed scheme, the
time of the occurrence is transmitted to the head node without waiting for a
synchronization message and the time translation is done at the head node. As
for the receiver clock skew compensation, the use of network time protocol (NTP)
\cite{mills91:_inter} was suggested in \cite{elson01:_time}; because local-clock
resolution and skew are minimized by the control-feedback design (e.g., based on
phase-locked loop (PLL)), however, the NTP is not proper for
low-power/complexity sensor nodes. In this regard the clock skew is handled by a
one-way, low-complexity asynchronous SCFR scheme described in \cite{Kim:13-1}
for the proposed scheme.

\textit{Reverse Two-Way Message Exchanges}: The ReversePTP proposed in
\cite{mizrahi14:_using_rever_softw_defin_networ} for time distribution in
software defined networks (SDNs) is also similar to our work because the
relationship and the direction of clock distribution between one master (i.e.,
the head node) and many slaves (i.e., sensor nodes) are reversed in both
schemes. Their contexts (i.e., SDNs vs WSNs), however, are quite different from
each other. As a consequence, the energy efficiency is not the main focus of the
ReversePTP. For example, the slaves in the ReversePTP distribute their times to
the master by periodically sending Sync messages. In the proposed scheme, it is
the head node that distributes its time (i.e., clock frequency); if there is no
data measurement, a sensor node does not transmit any message back to the head
node for time synchronization.

\textit{Passive Listening}: The idea of passive listening to timestamped
messages from other nodes is also discussed in
\cite{noh07:_pairw_broad_clock_synch_wirel_sensor_networ} and
\cite{chepuri13:_joint}. The receiver-only synchronization (ROS) scheme proposed
in \cite{noh07:_pairw_broad_clock_synch_wirel_sensor_networ} enables a subset of
sensor nodes to achieve synchronization by overhearing the conventional two-way
synchronization message exchanges of a pair of sensor nodes, called
\textit{super nodes}; because of the additional information from the overheard
two-way messages exchanges, the nodes can estimate both clock offset and clock
skew without any message transmission. The application of ROS scheme, however,
is limited to the sensor nodes located in the ``Region of Pairwise
Synchronization'' where the messages from both super nodes can reach. Also, two
super nodes performing two-way message exchanges are needed for ROS scheme.

The joint estimation algorithm based on prewhitening of observation models and
least squares (LS) in \cite{chepuri13:_joint} can be implemented either in a
centralized or distributed ways, but the distributed approach incurs additional
broadcast messages to distribute timestamps. In the proposed scheme, because the
clock skew and offset estimation procedures are separated from each other, a
distributed recursive algorithm can be applied for the clock skew estimation
based on passive listening without incurring additional message broadcasting.

\section{Conclusions}
\label{sec:conclusions}
\balance 
In this paper we have proposed an energy-efficient time synchronization scheme
for asymmetric WSNs, which is based on the asynchronous SCFR for one-way clock
skew estimation/compensation at sensor nodes and reverse two-way message
exchanges for clock offset estimation/translation at the head node in order to
minimize the number of message transmissions at sensor nodes. Taking notice of
the asymmetry between the head node and sensor nodes in terms of processing
power and supplied energy, we move most of time synchronization operations to
the head node in the proposed scheme and thereby reduce the complexity and power
consumption of sensor nodes for time synchronization.

As for the one-way clock skew estimation/compensation, i.e., the only major
operation to be done at sensor nodes, because the low complexity of the
algorithm is critical for power efficiency, we use the CR estimator proposed in
\cite{Kim:13-1} for asynchronous SCFR, which is an unbiased estimator attaining
the lower bound for a Gaussian delay as stated in
Proposition~\ref{th:ee_skew}. The comparative analysis presented in
Section~\ref{sec:performance-one-way} shows that the CR estimator can provide
relatively good performance with the lowest complexity among the estimators
considered irrespective of the delay distribution. The simulation results in
Section~\ref{sec:meas-time-estim} in fact verifies that the proposed time
synchronization scheme with the CR as the clock skew estimator provides the best
performance in terms of the accuracy of measurement time estimation and, as an
indirect measure of energy consumption, the number of message transmissions and
receptions at a sensor node. In addition, if there are no strict timing
requirements for the processing of measurement data, we can further reduce the
number of message transmissions by bundling measurement data in a
``Report/Response'' message without significantly affecting the time
synchronization performance, which is demonstrated by the simulation results in
Section~\ref{sec:effect-bundl-meas}.

Note that in this paper we have focused on the essential aspects of the proposed
time synchronization scheme, i.e., the one-way clock skew
estimation/compensation and the reverse two-way message exchanges combined with
measurement data report messages, which leaves room for further extensions of
the proposed scheme; examples include the operations with multiple head nodes
for redundancy and possibly localization as well \cite{chepuri13:_joint} and the
application of adaptive synchronization interval \cite{akhlaq13:_rtsp}.

Of possible extensions, we have already shown how the proposed scheme can be
extended to a hierarchical structure for network-wide, multi-hop synchronization
through either simple packet-relaying or time-translating gateway nodes. The
tradeoff between the complexity of gateway nodes and the performance of time
synchronization for the two approaches is an important topic for further
research and will be addressed in a followup work.

\appendices
\section{Proof of Proposition 1}
\label{sec:proof-proposition-1}
We first rephrase \eqref{eq:arrival_timestamp} as follows:
\begin{equation}
  \label{eq:arrival_timestamp_with_known_delay}
  t_{a,i}(k) = R_{i}t_{d}(k) + \theta_{i} + d + n(k) ,
\end{equation}
whee $d$ is a fixed portion of delay, which is assumed to be \textit{known}, and
$n(k)$ is a random portion of delay following a zero-mean white Gaussian
distribution with variance $\sigma^{2}$. From
\eqref{eq:arrival_timestamp_with_known_delay}, we can construct a likelihood
function and a log-likelihood function for the collective observation
$\bm{t_{a,i}}{\triangleq}\left\{t_{a,i}(j)\right\}^{k}_{j=0}$ and the parameters
$\theta_{i}$ and $R_{i}$ as follows:
\begin{align}
  \label{eq:lf_mle_skew}
  \MoveEqLeft \mathcal{L}(\theta_{i},R_{i};\bm{t_{a,i}}) = \notag \\
& \prod^{k}_{j=1}\dfrac{1}{\sqrt{2\pi\sigma^{2}}}
  \exp\left(-\dfrac{\left(t_{a,i}(j)-\left(R_{i}t_{d}(j)+\theta_{i}+d\right)\right)^{2}}{2\sigma^{2}}\right) ,
\end{align}
and
\begin{align}
  \label{eq:llf_mle_skew}
  \MoveEqLeft \ln \mathcal{L}(\theta_{i},R_{i};\bm{t_{a,i}}) = -\dfrac{k}{2}\ln\left(2\pi\sigma^{2}\right) \notag \\
&-\dfrac{1}{2\sigma^{2}} \sum^{k}_{j=0} \left\{t_{a,i}(j)-\left(R_{i}t_{d}(j)+\theta_{i}+d\right)\right\}^{2} .
\end{align}
Then the first derivatives of the log likelihood function with respect to
$\theta_{i}$ and $R_{i}$ are given by
\begin{align}
  \label{eq:1st_pd_llf_mle_offset}
  \MoveEqLeft \dfrac{\partial \ln\mathcal{L}(\theta_{i},R_{i};\bm{t_{a,i}})}{\partial \theta_{i}} = \notag \\
& \dfrac{1}{\sigma^{2}}
  \sum^{k}_{j=0}\left\{t_{a,i}(j)-\left(R_{i}t_{d}(j)+\theta_{i}+d\right)\right\} ,
\end{align}
\begin{align}
  \label{eq:1st_pd_llf_mle_skew}
  \MoveEqLeft \dfrac{\partial \ln
  \mathcal{L}(\theta_{i},R_{i};\bm{t_{a,i}})}{\partial R_{i}} = \notag \\
& \dfrac{1}{\sigma^{2}}
  \sum^{k}_{j=0}t_{d}(j)\left\{t_{a,i}(j)-\left(R_{i}t_{d}(j)+\theta_{i}+d\right)\right\} .
\end{align}
From \eqref{eq:1st_pd_llf_mle_offset} and \eqref{eq:1st_pd_llf_mle_skew},
therefore, we obtain the joint one-way MLE of clock offset
$\hat{\theta}^{ML}_{i}(k)$ and skew $\hat{R}^{ML}_{i}(k)$ as follows:
\begin{equation}
  \label{eq:joint_mle_offset}
  \hat{\theta}^{ML}_{i}(k) = \overline{t_{a,i}} - \hat{R}^{ML}_{i}(k) \overline{t_{d}}
  - d ,
\end{equation}
\begin{equation}
  \label{eq:joint_mle_skew}
  \hat{R}^{ML}_{i}(k) = \dfrac{
    \sum^{k}_{j=0}\left\{t_{a,i}(j)-\left(\hat{\theta}^{ML}_{i}(k)+d\right)\right\}
  }{
    \sum^{k}_{j=0}t^{2}_{d}(j)
  } .
\end{equation}
After some manipulations, we can obtain closed-form expressions for
$\hat{\theta}^{ML}_{i}(k)$ and $\hat{R}^{ML}_{i}(k)$ given in
\eqref{eq:joint_mle_offset_closed_form} and
\eqref{eq:joint_mle_skew_closed_form}, respectively.

To obtain the CRLBs, we need the second derivatives of the log likelihood
function with respect to $\theta_{i}$ and $R_{i}$, which are given by
\begin{equation}
  \label{eq:2nd_pd_llf_mle_offset}
  \dfrac{
    \partial^{2} \ln \mathcal{L}(\theta_{i},R_{i};\bm{t_{a,i}})
  }{
    \partial \theta_{i}^{2}
  }
  = -\dfrac{k}{\sigma^{2}} ,
\end{equation}
\begin{equation}
  \label{eq:2nd_pd_llf_mle_skew}
  \dfrac{
    \partial^{2} \ln \mathcal{L}(\theta_{i},R_{i};\bm{t_{a,i}})
  }{
    \partial R_{i}^{2}
  }
  = -\dfrac{\sum^{k}_{j=0}t_{d}(j)^{2}}{\sigma^{2}} 
  = -\dfrac{k}{\sigma^{2}} \overline{t_{d}^{2}} ,
\end{equation}
\begin{equation}
  \label{eq:2nd_pd_llf_mle_offset_skew}
  \dfrac{
    \partial^{2} \ln \mathcal{L}(\theta_{i},R_{i};\bm{t_{a,i}})
  }{
    \partial \theta_{i} \partial R_{i}
  }
  = -\dfrac{\sum^{k}_{j=0}t_{d}(j)}{\sigma^{2}}
  = -\dfrac{k}{\sigma^{2}} \overline{t_{d}} .
\end{equation}
Note that there is no need to take expectations because there are no terms
related with $t_{a,i}(j)$. The Fisher information matrix in this case becomes
\begin{align}
  \label{eq:joint_mle_fisher}
  \arraycolsep=4pt\def\arraystretch{1.5}
  \bm{\mathrm{I}}(\theta_{i},R_{i}) = \dfrac{k}{\sigma^{2}}
  \left[
  \begin{array}{cc}
    1 & \overline{t_{d}} \\
    \overline{t_{d}} & \overline{t_{d}^{2}}
  \end{array}
                       \right] .
\end{align}
Because the CRLBs are the diagonal elements of the inverse of the Fisher
information matrix \cite[p.~40]{kay93:_fundam}, we obtain
\begin{align}
  \label{eq:joint_mle_fisher_inverse}
  \arraycolsep=1.5pt\def\arraystretch{1.5}
  \bm{\mathrm{I}}^{-1}(\theta_{i},R_{i}) = \dfrac{\sigma^{2}}{k\left\{\overline{t_{d}^{2}}-\left(\overline{t_{d}}\right)^{2}\right\}}
  \left[
  \begin{array}{cc}
    \overline{t_{d}^{2}} & -\overline{t_{d}} \\
    -\overline{t_{d}} & 1
  \end{array}
                        \right] .
\end{align}
From \eqref{eq:joint_mle_fisher_inverse}, therefore, we can obtain the CRLBs of
clock offset and skew for the Gaussian delay model given in
\eqref{eq:joint_mle_offset_crlb} and \eqref{eq:joint_mle_skew_crlb}, i.e.,
\[
\operatorname{Var}\left(\hat{\theta}_{i}(k)\right) \geq \dfrac{ \sigma^{2} \cdot
  \overline{t_{d}^{2}} }{
  k\left\{\overline{t_{d}^{2}}-\left(\overline{t_{d}}\right)^{2}\right\} } ,
\]
\[
\operatorname{Var}\left(\hat{R}_{i}(k)\right) \geq \dfrac{ \sigma^{2} }{
  k\left\{\overline{t_{d}^{2}}-\left(\overline{t_{d}}\right)^{2}\right\} } .
\]

Note that the joint MLEs of clock offset $\hat{\theta}^{ML}_{i}(k)$ and skew
$\hat{R}^{ML}_{i}(k)$ are \textit{efficient estimators}: First, from
\eqref{eq:joint_mle_skew_closed_form}, we obtain
\begin{align}
  \label{eq:joint_mle_skew_mean}
  \operatorname{E}\left[\hat{R}^{ML}_{i}(k)\right] = \dfrac{
  \operatorname{E}\left[\overline{t_{d}t_{a,i}}\right] -
  \overline{t_{d}} \operatorname{E}\left[\overline{t_{a,i}}\right]
  }{
  \overline{t^{2}_{d}} - \left(\overline{t_{d}} \right)^{2}
  } .
\end{align}
From \eqref{eq:arrival_timestamp_with_known_delay}, we have
\begin{align}
  \label{eq:td_ta_mean}
  \operatorname{E}\left[\overline{t_{d}t_{a,i}}\right]
  = R_{i}\overline{t_{d}^{2}} + \left(\theta_{i} + d\right)\overline{t_{d}} ,
\end{align}
\begin{align}
  \label{eq:ta_mean}
  \operatorname{E}\left[\overline{t_{a,i}}\right]
  = R_{i}\overline{t_{d}} + \left(\theta_{i} + d\right) .
\end{align}
Inserting \eqref{eq:td_ta_mean} and \eqref{eq:ta_mean} into
\eqref{eq:joint_mle_skew_mean}, we obtain
$\operatorname{E}\left[\hat{R}^{ML}_{i}(k)\right]{=}R_{i}$, which shows that
$\hat{R}^{ML}_{i}(k)$ is an unbiased estimator.

Secondly, from \eqref{eq:joint_mle_skew_closed_form}, we also obtain
\begin{align}
  \label{eq:joint_mle_skew_variance}
  \operatorname{Var}\left(\hat{R}^{ML}_{i}(k)\right) = \dfrac{
  \operatorname{Var}\left(\overline{t_{d}t_{a,i}}\right) -
  \left(\overline{t_{d}}\right)^{2} \operatorname{Var}\left(\overline{t_{a,i}}\right)
  }{
  \left\{\overline{t^{2}_{d}} - \left(\overline{t_{d}} \right)^{2}\right\}^{2}
  } .
\end{align}
From \eqref{eq:arrival_timestamp_with_known_delay}, we have
\begin{align}
  \label{eq:td_ta_variance}
  \operatorname{Var}\left(\overline{t_{d}t_{a,i}}\right)
      = \dfrac{
      \sum_{j=0}^{k}t_{d}(j)^{2}\operatorname{Var}\left(n(k)\right)
      }{
      k^{2}
      } = \dfrac{\sigma^{2}}{k} \overline{t_{d}^{2}} ,
\end{align}
\begin{align}
  \label{eq:ta_variance}
  \operatorname{Var}\left(\overline{t_{a,i}}\right)
  = \dfrac{
  \sum_{j=0}^{k}\sigma^{2}
  }{
  k^{2}
  } = \dfrac{\sigma^{2}}{k} .
\end{align}
Inserting \eqref{eq:td_ta_variance} and \eqref{eq:ta_variance} into
\eqref{eq:joint_mle_skew_variance}, we obtain
\[
\operatorname{Var}\left(\hat{R}^{ML}_{i}(k)\right) = \dfrac{ \sigma^{2} }{
  k\left\{\overline{t_{d}^{2}}-\left(\overline{t_{d}}\right)^{2}\right\} } ,
\]
which is the CRLB in \eqref{eq:joint_mle_skew_crlb}. Similar procedures can be
taken for $\hat{\theta}^{ML}_{i}(k)$ to show that it is also an efficient
estimator. \hfill\IEEEQED

\section{Proof of Proposition 2}
\label{sec:proof-proposition-2}
It is straightforward to see that $\hat{R}^{CR}_{i}(k)$ is unbiased because
\begin{equation}
  \label{eq:cr_mean}
  \operatorname{E}\left[\hat{R}^{CR}_{i}(k)\right] = \operatorname{E}\left[R_{i} +
    \dfrac{1}{\tilde{t}_{d}(k)}\tilde{d}(k)\right] = R_{i} .
\end{equation}
As for the variance, we first construct from \eqref{eq:rot_model} a likelihood
function for the observation $\tilde{t}_{a,i}(k)$ and the parameter $R_{i}$:
\begin{align}
  \label{eq:lf_cr_skew}
  \mathcal{L}(R_{i};\tilde{t}_{a,i}(k)) = 
  \dfrac{1}{\sqrt{4\pi\sigma^{2}}}
  \exp\left(-\dfrac{\left(\tilde{t}_{a,i}(k)-R_{i}\tilde{t}_{d}(k)\right)^{2}}{4\sigma^{2}}\right) .
\end{align}
Then the first and the second derivatives of the log likelihood function are
given by
\begin{align}
  \label{eq:1st_pd_llf_cr_skew}
  \dfrac{\partial \ln \mathcal{L}(R_{i};\tilde{t}_{a,i}(k))}{\partial R_{i}}
      = \dfrac{\tilde{t}_{a,i}(k)-R_{i}\tilde{t}_{d}(k)}{2\sigma^{2}} \tilde{t}_{d}(k) ,
\end{align}
\begin{equation}
  \label{eq:2nd_pd_llf_cr_skew}
  \dfrac{\partial^{2} \ln \mathcal{L}(R_{i};\tilde{t}_{a,i}(k))}{\partial^{2} R_{i}}
  = -\dfrac{\tilde{t}_{d}(k)^{2}}{2\sigma^{2}} .
\end{equation}
From \eqref{eq:1st_pd_llf_cr_skew}, we can see that $\hat{R}^{CR}_{i}(k)$ is in
fact the estimator maximizing the log-likelihood function. Also from
\eqref{eq:2nd_pd_llf_cr_skew}, applying the same steps to derive a CRLB, we can
obtain the lower bound of clock skew as follows
\begin{equation}
  \label{eq:lb_cr_skew}
  \operatorname{Var}\left(\hat{R}^{CR}_{i}(k)\right) \geq
  \dfrac{2\sigma^{2}}{\tilde{t}_{d}(k)^{2}} .
\end{equation}
Because the variance of $\hat{R}^{CR}_{i}(k)$ is given by
\begin{equation}
  \label{eq:cr_variance}
  \operatorname{Var}\left(\hat{R}^{CR}_{i}(k)\right) = 
  \operatorname{Var}\left(R_{i}+\dfrac{\tilde{d}(k)}{\tilde{t}_{d}(k)}\right)  =
  \dfrac{2\sigma^{2}}{\tilde{t}_{d}(k)^{2}} ,
\end{equation}
$\hat{R}^{CR}_{i}(k)$ attains the lower bound given by \eqref{eq:lb_cr_skew}.
\hfill\IEEEQED


\begin{thebibliography}{10}
\providecommand{\url}[1]{#1}
\csname url@samestyle\endcsname
\providecommand{\newblock}{\relax}
\providecommand{\bibinfo}[2]{#2}
\providecommand{\BIBentrySTDinterwordspacing}{\spaceskip=0pt\relax}
\providecommand{\BIBentryALTinterwordstretchfactor}{4}
\providecommand{\BIBentryALTinterwordspacing}{\spaceskip=\fontdimen2\font plus
\BIBentryALTinterwordstretchfactor\fontdimen3\font minus
  \fontdimen4\font\relax}
\providecommand{\BIBforeignlanguage}[2]{{%
\expandafter\ifx\csname l@#1\endcsname\relax
\typeout{** WARNING: IEEEtran.bst: No hyphenation pattern has been}%
\typeout{** loaded for the language `#1'. Using the pattern for}%
\typeout{** the default language instead.}%
\else
\language=\csname l@#1\endcsname
\fi
#2}}
\providecommand{\BIBdecl}{\relax}
\BIBdecl

\bibitem{wu11:_clock_synch_wirel_sensor_networ}
Y.-C. Wu, Q.~Chaudhari, and E.~Serpedin, ``Clock synchronization of wireless
  sensor networks,'' \emph{{IEEE} Signal Process. Mag.}, vol.~28, no.~1, pp.
  124--138, 2011.

\bibitem{akhlaq13:_rtsp}
M.~Akhlaq and T.~R. Sheltami, ``{RTSP}: An accurate and energy-efficient
  protocol for clock synchronization in {WSNs},'' \emph{{IEEE} Trans. Instrum.
  Meas.}, vol.~62, no.~3, pp. 578--589, Mar. 2013.

\bibitem{macii09:_power_wirel_sensor_networ}
D.~Macii, A.~Ageev, and A.~Somov, ``Power consumption reduction in wireless
  sensor networks through optimal synchronization,'' in \emph{Proc. I2MTC '09},
  May 2009, pp. 1346--1351.

\bibitem{elson02:_fine}
J.~Elson, L.~Girod, and D.~Estrin, ``Fine-grained network time synchronization
  using reference broadcasts,'' in \emph{Proc. of 5th Symp. Operating System
  Design and Implementation}, Dec. 2002, pp. 147--163.

\bibitem{trump01:_maxim}
T.~Trump, ``Maximum likelihood trend estimation in exponential noise,''
  \emph{{IEEE} Trans. Signal Process.}, vol.~49, no.~9, pp. 2087--2095, Sep.
  2001.

\bibitem{rajan11:_joint}
R.~T. Rajan and A.-J. van~der Veen, ``Joint ranging and clock synchronization
  for a wireless network,'' in \emph{Proc. of {CAMSAP} 2011}, Dec. 2011, pp.
  297--300.

\bibitem{chepuri13:_joint}
S.~P. Chepuri, R.~T. Rajan, G.~Leus, and A.-J. van~der Veen, ``Joint clock
  synchronization and ranging: Asymmetrical time-stamping and passive
  listening,'' \emph{{IEEE} Signal Process. Lett.}, vol.~20, no.~1, pp. 51--54,
  Jan. 2013.

\bibitem{Moon:99}
S.~B. Moon, P.~Skelly, and D.~Towsley, ``Estimation and removal of clock skew
  from network delay measurements,'' in \emph{Proc. 1999 IEEE INFOCOM}, New
  York, NY, Mar. 1999.

\bibitem{Kim:13-1}
K.~S. Kim, ``Asynchronous source clock frequency recovery through aperiodic
  packet streams,'' \emph{{IEEE} Commun. Lett.}, vol.~17, no.~7, pp.
  1455--1458, Jul. 2013.

\bibitem{mainwaring02:_wirel_sensor_networ_habit_monit}
A.~Mainwaring, J.~P.~R. Szewczyk, D.~Culler, and J.~Anderson, ``Wireless sensor
  networks for habitat monitoring,'' in \emph{Proc. WSNA '02}.\hskip 1em plus
  0.5em minus 0.4em\relax New York, NY, USA: ACM, 2002, pp. 88--97.

\bibitem{mills91:_inter}
D.~L. Mills, ``Internet time synchronization: The network time protocol,''
  \emph{{IEEE} Trans. Commun.}, vol.~39, no.~10, pp. 1482--1493, Oct. 1991.

\bibitem{ganeriwal03:_timin_protoc_sensor_networ}
S.~Ganeriwal, R.~Kumar, and M.~B. Srivastava, ``Timing-sync protocol for sensor
  networks,'' in \emph{Proc. {SenSys}'03}.\hskip 1em plus 0.5em minus
  0.4em\relax New York, NY, USA: ACM, 2003, pp. 138--149.

\bibitem{Kim:14-4}
K.~S. Kim, ``Comments on "ieee 1588 clock synchronization using dual slave
  clocks in a slave",'' \emph{{IEEE} Commun. Lett.}, vol.~18, no.~6, pp.
  981--982, Jun. 2014.

\bibitem{noh07:_novel}
K.-L. Noh, Q.~M. Chaudhari, E.~Serpedin, and B.~W. Suter, ``Novel clock phase
  offset and skew estimation using two-way timing message exchanges for
  wireless sensor networks,'' \emph{{IEEE} Trans. Commun.}, vol.~55, no.~4, pp.
  766--777, Apr. 2007.

\bibitem{guchhait15:_joint_minim_varian_unbias_maxim}
A.~Guchhait and R.~M. Karthik, ``Joint minimum variance unbiased and maximum
  likelihood estimation of clock offset and skew in one-way packet
  transmission,'' in \emph{Proc. {IEEE} 81st Vehicular Technology Conference
  ({VTC2015}-Spring)}, May 2015, pp. 1--6.

\bibitem{kay93:_fundam}
S.~M. Kay, \emph{Fundamentals of statistical signal processing: Estimation
  theory}.\hskip 1em plus 0.5em minus 0.4em\relax Upper Saddle River, NJ, USA:
  Prentice-Hall, Inc., 1993, vol.~1.

\bibitem{elson01:_time}
J.~Elson and D.~Estrin, ``Time synchronization for wireless sensor networks,''
  in \emph{Proc. {IPDPS} 2001}, San Francisco, CA, USA, Apr. 2001.

\bibitem{mizrahi14:_using_rever_softw_defin_networ}
T.~Mizrahi and Y.~Moses, ``Using reverseptp to distribute time in software
  defined networks,'' in \emph{Proc. of {ISPCS} 2014}, Sep. 2014, pp. 112--117.

\bibitem{noh07:_pairw_broad_clock_synch_wirel_sensor_networ}
N.~Kyoung-lae and E.~Serpedin, ``Pairwise broadcast clock synchronization for
  wireless sensor networks,'' in \emph{Proc. {WoWMoM} 2007}, Jun. 2007, pp.
  1--6.

\end{thebibliography}
\end{document}